\documentclass[amsmath,amssymb,aps,prd,twocolumn,superscriptaddress,floatfix,nopacs,preprintnumbers, nolongbibliography,nofootinbib] {revtex4-2}

\usepackage[utf8]{inputenc}
\usepackage{subfigure}
\usepackage[normalem]{ulem}
\usepackage[dvipsnames,table]{xcolor}
\usepackage{mathrsfs}
\usepackage{amssymb,bm}
\usepackage{dsfont}
\usepackage{amsmath}
\usepackage{mathtools}
\usepackage{slashed}
\usepackage{xfrac}
\usepackage{graphics}
\usepackage{graphicx}
\usepackage{physics}
\usepackage[h]{esvect}
\usepackage{accents}
\usepackage{standalone} 
\usepackage{graphicx}
\usepackage{tikz}
\usepackage{verbatim}
\standaloneconfig{mode=buildnew}
\usepackage{tcolorbox}

\definecolor{lcolor}{rgb}{0.5,0,0}
\definecolor{citcolor}{rgb}{0,0.3,0.0}

\renewcommand{\tr}[1]{\mathrm{Tr}\left[#1\right]}
\renewcommand{\Re}{\mathrm{Re}}

\renewcommand{\dd}{\mathrm{d}}
\newcommand{\DD}{\mathrm{D}}
\newcommand{\GG}{\mathcal{G}}
\newcommand{\be}{\begin{equation}}
\newcommand{\ee}{\end{equation}}
\newcommand{\half}{\frac{1}{2}}
\renewcommand{\bra}{\langle}
\renewcommand{\ket}{\rangle}

\usepackage{xargs}
\usepackage[colorinlistoftodos,prependcaption,textsize=footnotesize]{todonotes}
\newcommandx{\DM}[2][1=]{\todo[linecolor=orange,backgroundcolor=orange!25,bordercolor=orange,#1]{DM: #2}}
\newcommandx{\AI}[2][1=]{\todo[linecolor=magenta,backgroundcolor=magenta!25,bordercolor=magenta,#1]{AI: #2}}
\newcommandx{\TR}[2][1=]{\todo[linecolor=magenta,backgroundcolor=yellow!25,bordercolor=yellow,#1]{TR: #2}}

\usepackage[breaklinks,colorlinks,citecolor=citcolor,urlcolor=blue,linkcolor=lcolor]{hyperref}

\newcommand{\nn}{\nonumber}
\begin{document}

\preprint{RIKEN-iTHEMS-Report-26}

\author{Gert Aarts}
\email{gert.aarts@uni-bielefeld.de}
\affiliation{Faculty of Physics, Bielefeld University,
D-33615 Bielefeld, Germany}

\author{Diaa E.~Habibi}
\email{n.e.habibi@swansea.ac.uk}
\affiliation{Centre for Quantum Fields and Gravity, Department of Physics, Swansea University, Swansea SA2 8PP, United Kingdom}

\author{Andreas Ipp}
\email{ipp@hep.itp.tuwien.ac.at}
\affiliation{Institute for Theoretical Physics, TU Wien, Wiedner Hauptstraße  8-10/136, A-1040 Vienna, Austria}

\author{David I.~M\"{u}ller}
\email{dmueller@hep.itp.tuwien.ac.at}
\affiliation{Institute for Theoretical Physics, TU Wien, Wiedner Hauptstraße  8-10/136, A-1040 Vienna, Austria}

\author{Thomas R.~Ranner}
\email{thomas.ranner@tuwien.ac.at}
\affiliation{Institute for Theoretical Physics, TU Wien, Wiedner Hauptstraße  8-10/136, A-1040 Vienna, Austria}

\author{Lingxiao Wang}
\email{lingxiao.wang@riken.jp}
\affiliation{RIKEN Center for Interdisciplinary Theoretical and Mathematical Sciences (iTHEMS), Wako, Saitama 351-0198, Japan}
\affiliation{Institute for Physics of Intelligence, Graduate School of Science, The University of Tokyo, Bunkyo-ku, Tokyo 113-0033, Japan}

\author{Wei Wang}
\email{wei.wang@sjtu.edu.cn}
\affiliation{State Key Laboratory of Dark Matter Physics, Key Laboratory for Particle Astrophysics and Cosmology (MOE),
Shanghai Key Laboratory for Particle Physics and Cosmology,
Shanghai Jiao Tong University, Shanghai 200240, China}
\affiliation{Southern Center for Nuclear-Science Theory (SCNT), Institute of Modern Physics, Chinese Academy of Sciences, 96 South Sihuan Rd. Huicheng District, Huizhou 516000, Guangdong, China}

\author{Qianteng Zhu}
\email{zhuqianteng@sjtu.edu.cn}
\affiliation{RIKEN Center for Interdisciplinary Theoretical and Mathematical Sciences (iTHEMS), Wako, Saitama 351-0198, Japan}
\affiliation{State Key Laboratory of Dark Matter Physics, Key Laboratory for Particle Astrophysics and Cosmology (MOE),
Shanghai Key Laboratory for Particle Physics and Cosmology,
Shanghai Jiao Tong University, Shanghai 200240, China}

\title{Generalizable Equivariant Diffusion Models for Non-Abelian Lattice Gauge Theory: \\
Analytical Foundations and Numerical Scaling Tests}

\begin{abstract}
Gauge-equivariant diffusion models can be used to generate ensembles of field configurations in non-abelian lattice gauge theories.  
Previously, we demonstrated the feasibility of modeling these theories by combining L-CNN-based diffusion models with the Metropolis-adjusted annealed Langevin algorithm (MAALA). 
Here we focus on the underlying analytical framework. In particular, we present a rigorous derivation of the reverse-time stochastic differential equation on the group manifold, discuss the gauge invariance of the denoising loss function, and provide extensive numerical benchmarks regarding the scaling behavior with respect to the lattice volume and the inverse coupling $\beta$. Full analytical expressions for arbitrarily sized Wilson loops and the topological charge susceptibility during the forward process are given as well.

\end{abstract}

\maketitle 

\section{Introduction}

The lattice approximation is the most successful computational tool for analyzing the non-perturbative aspects of quantum chromodynamics (QCD) \cite{FlavourLatticeAveragingGroupFLAG:2024oxs,Aarts:2023vsf}. Many static and thermodynamic observables, such as mass spectra, topological properties, pressure and energy density, are beyond the reach of traditional perturbation theory and can generally only be studied within the lattice regularization. The main idea is to replace continuous space-time, on which gluonic and fermionic fields are defined, by a discrete, generally hypercubic, lattice. As a result, intractable path integrals in the continuum become finite (but high) dimensional integrals that can be solved numerically using Markov chain Monte Carlo (MCMC) simulations.

In order to obtain results for the continuum theory, the non-zero lattice spacing $a$ must eventually be sent to zero, with the aim to remove unphysical lattice artifacts. In lattice QCD, the spacing $a$ is not a directly controllable parameter. Instead, $a$ is determined through the inverse coupling $\beta=2N_c/g_{\rm YM}^2$ where $g_{\rm YM}$ is the Yang-Mills coupling constant. The $a\rightarrow 0$ ($\beta \rightarrow \infty$) limit can only be approached using extrapolations, i.e., by simulating at a series of different finite lattice spacings and inferring the continuum value ($a=0$) through statistical fits. At fixed physical volumes, approaching the continuum limit and reducing statistical Monte Carlo errors to the desired precision require increasingly fine lattices and substantial computational effort. Additionally, taking the continuum limit is severely hindered by \emph{critical slowing down} and \emph{topological freezing} \cite{Schaefer:2010hu, Alles:1996vn,DelDebbio:2004xh}. 
Traditional Markov chain methods exhibit a power-law increase in autocorrelation times for all observables as the lattice spacing decreases. At smaller lattice spacings, topological properties of QCD become almost adversarial to the Markov chain approach: tunneling between topological sectors becomes exponentially rare, causing an additional exponential increase in autocorrelation times for topologically sensitive observables such as the topological susceptibility.
Therefore, the computational cost of generating statistically independent Monte Carlo samples and hence determining physical observables at a given statistical precision increases dramatically. This problem, therefore, presents two different lines of attack: reducing lattice artifacts at coarse lattice spacings or decreasing autocorrelation times for simulations on fine lattices. In both cases, algorithmic improvements and, more recently, machine learning (ML) methods, rigorously applied to lattice field theory, have made great progress over the past decades, see e.g., Refs.~\cite{Cranmer:2023xbe,Aarts:2025gyp,Tomiya:2025quf} for reviews with a focus on machine learning.  
Here we briefly review several approaches aimed at addressing issues associated with the continuum limit. 

Improving the lattice approximation goes back to Symanzik \cite{Symanzik:1983dc, Symanzik:1983gh}. The idea is to use more complicated lattice actions and observables 
in order to reduce lattice artifacts at a given lattice spacing $a$. With such improved lattice simulations, accurate continuum extrapolations can be obtained without needing very fine lattice spacings. Improved actions for lattice gauge theory have been worked out in detail (see, e.g., Refs.~\cite{Weisz:1982zw,Weisz:1983bn,Luscher:1984xn}).
The pinnacle of the improvement program is to use perfect lattice actions \cite{Hasenfratz:1993sp, DeGrand:1995ji}, in which lattice artifacts are either strongly reduced (classically perfect fixed-point actions) or entirely eliminated (quantum perfect actions). This approach follows the ideas of the Wilsonian renormalization group (RG) on the lattice \cite{Wilson:1973jj, Wilson:1974mb, Bell:1975wtp}. Although perfect lattice actions are local, they require truncations to be useful in practice. Even these truncations are very complicated functions of the fields and thus costly to simulate. However, recently, the fixed-point improvement program has been reinvigorated through ML: machine-learned classically perfect actions have shown remarkably good scaling toward the continuum in practical Monte Carlo simulations for four-dimensional SU(3) theory \cite{Holland:2024muu, Holland:2025fsa}.

Alternatively, improvements to Monte Carlo algorithms can counteract or circumvent either critical slowing down or topological freezing by other means: for example, topological freezing can be alleviated using 
open boundary conditions \cite{Luscher:2011kk}, parallel tempering \cite{Hasenbusch:2017unr, Bonanno:2020hht, Bonanno:2024zyn}, or topology-changing updates \cite{Eichhorn:2022wxn} and metadynamics \cite{Eichhorn:2023uge, Eichhorn:2026eok}. More generally, critical slowing down, which causes large autocorrelation times of infrared modes, is reduced using Fourier acceleration \cite{Batrouni:1985jn, Duane:1988vr} and Riemannian Manifold Hybrid MC (RMHMC) \cite{Nguyen:2021zgx, Christ:2024ydh, Jung:2024nuv}, multiscale thermalization \cite{Endres:2015yca, Detmold:2018zgk} or upscaling algorithms \cite{Hasenfratz:2026azk}.
One particularly interesting direction due to Lüscher \cite{Luscher:2009eq} is the trivializing map. The idea is to find a differentiable, bijective map from an ensemble of configurations in the strong-coupling limit ($\beta = 0$) to an 
ensemble at finite $\beta$. Since the former is easy to sample without autocorrelation, the trivializing map would allow sampling at finite $\beta$ without encountering critical slowing down. Similar to perfect actions, trivializing maps are difficult to construct, and the computational advantage over standard Monte Carlo methods might be slim \cite{Engel:2011re}. However, ML techniques can also be very beneficial here. Machine-learned trivializing maps have recently proven successful in two-dimensional scalar field theory \cite{Albandea:2023wgd} and gauge theory \cite{Bacchio:2022vje, Gerdes:2024rjk}.

A closely related idea originating in ML is normalizing flows (NFs) \cite{rezende2015variational, dinh2016density} (see also Ref.~\cite{papamakarios2021normalizing} for an historic overview): a complicated probability distribution, the ``target'', is mapped to a simple normal distribution (hence the term ``normalizing flow''). Similar to the trivializing map, NFs are invertible and differentiable and ideally exhibit no autocorrelations among samples. Thus, NFs in principle solve both critical slowing down and, in the case of theories with non-trivial topology, also topological freezing. 
The main difference between the trivializing map and the NF is that trivializing maps are constructed analytically, e.g., perturbatively or via the renormalization Group, while NFs are trained numerically via gradient descent, minimizing the Kullback-Leibler divergence. 
Both use the known action as the target.
NFs can also be formulated as continuous systems using ordinary differential equations (ODEs), resulting in continuous flows or CNFs \cite{chen2018neural}. In lattice field theory, NFs have been studied extensively across many different theories: by now there are studies of NFs for two-dimensional scalar fields \cite{Albergo:2019eim, DelDebbio:2021qwf}, two-dimensional \cite{Kanwar:2020xzo,Boyda:2020hsi} and four-dimensional \cite{Abbott:2023thq} gauge theory, fermionic theories \cite{Albergo:2021bna, Albergo:2022qfi, Abbott:2022zhs}, and also four-dimensional QCD, albeit on a small volume and at relatively strong coupling \cite{Abbott:2022hkm}; CNFs have been studied in scalar field theory \cite{deHaan:2021erb, Gerdes:2022eve}. It was demonstrated that NFs effectively mitigate topological freezing, but scaling to large four-dimensional lattices with small lattice spacing has proven difficult in practice \cite{Abbott:2022zsh}. Beyond direct sampling, NFs are also very effective at variance reduction \cite{Abbott:2024kfc, Abbott:2026ylv}.

Stochastic NFs (SNFs) \cite{NEURIPS2020_41d80bfc, Caselle:2022acb} combine non-equilibrium MCMC (NE-MCMC) simulations (for lattice applications, see e.g.~\cite{Caselle:2016wsw, Caselle:2018kap, Francesconi:2020fgi, Bulgarelli:2023ofi, Bonanno:2024udh})
with intermediate NF maps. The NE-MCMC approach is based on a sequence of MC simulations with predefined probability densities (a \textit{protocol}), which ends with the desired target distribution. 
By virtue of reweighting and Jarzynski's identity \cite{Jarzynski:1996oqb,Caselle:2016wsw}, any remaining bias can be removed, which renders NE-MCMC exact. 
SNFs extend the NE-MCMC approach by introducing NFs as intermediate flow maps between steps, thereby improving sampling efficiency. These intermediate NFs are easier to train compared to direct sampling due to overlapping probability densities of consecutive steps. 
NE-MCMC simulations, and similarly SNFs, have shown much better scaling than plain NFs with increasing lattice size \cite{Bulgarelli:2024brv}. Applications range from mitigating topological freezing in lattice gauge theory \cite{Bonanno:2025pdp, Bonanno:2026tle}, to reweighting pure-gauge SU(3) ensembles in four dimensions \cite{Bulgarelli:2024cqc}.

Diffusion models (DMs) \cite{sohldickstein2015deepunsupervisedlearningusing, song2020generativemodelingestimatinggradients, ho2020denoisingdiffusionprobabilisticmodels, song2021scorebasedgenerativemodelingstochastic} also connect a simple prior to a target distribution, but via a stochastic process rather than an invertible map. Given a training dataset and a specific diffusion process, machine learning models can be trained to reverse it, thereby enabling sampling. Since the forward and backward processes are formulated in terms of stochastic differential equations (SDEs), the method is quite accessible to physicists. 
In the context of lattice field theory, it was observed that DMs resemble stochastic quantization \cite{Wang:2023exq, Fukushima:2024oij}, or, more closely, stochastic RG \cite{Carosso:2019qpb, Cotler:2023lem}. Moreover, SDE-based DMs allow for multiple sampling methods: either through the reverse SDE, or equivalently, the reverse ODE, which enables deterministic sampling and reweighting \cite{song2021scorebasedgenerativemodelingstochastic}. Similar to NFs, sampled configurations are independent and thus uncorrelated. 
As with NFs, diffusion models have been applied to 
$\phi^4$ theory \cite{Cotler:2023lem,Wang:2023exq,Wang:2023sry,Tan:2026ibs, Sexty:2026aym, Kanwar:2026prt}, 
the XY model \cite{Tomiya:2026tbc}, 
the Schwinger model \cite{Vega:2026lrm},
the O(3) non-linear sigma model \cite{Kanwar:2026prt, Schubring:2026vnl}
and to data generated using complex Langevin dynamics \cite{Aarts:2025lpi}. 
More recently, the focus has also shifted to Abelian  \cite{Zhu:2025pmw,Vega:2025hgz} 
and non-abelian lattice gauge theories \cite{Aarts:2026zzr,Vega:2025xma, Alharazin:2026lcb, Komijani:2026lan}.
Finally, selected other approaches include physics-informed kernels (PIKs) \cite{Kapust:2025ida, Ihssen:2025hyl,Ihssen:2026njd}, the stochastic path sampler (SPS) \cite{Chen:2026bqw}, and optimal control \cite{Wang:2026pmv}.

To summarize, the current three main machine-learning-assisted approaches to sampling lattice field theories, which we refer to as neural lattice samplers below, are NFs (including CNFs and ML-based trivializing maps), SNFs, and DMs. When designing neural lattice samplers that can genuinely compete with traditional methods, it is prudent to adhere to a few principles, namely \emph{exact symmetries} (global and local), \emph{generalizability} across lattice sizes and couplings, and \emph{exactness}, that is, some kind of guarantee that the generated samples are representative of the target distribution. 
Starting with symmetries, in recent years it has proven indispensable to correctly implement the symmetries of the physical theory in the model architecture. Here we distinguish between the global lattice symmetries (translations, rotations, flips), global internal symmetries (e.g., $\mathbb Z_2$ in the case of $\phi^4$-theory), and local gauge symmetries. Translational symmetry on the lattice is rather easy to realize \cite{Bulusu:2021rqz} and is implemented in essentially all neural lattice samplers. However, exact rotation and mirror symmetry are notably absent from most works, or being realized only approximately. This might appear surprising, given that group-equivariant neural network architectures (e.g., G-CNNs or steerable CNNs) have been an active topic of research for the past decade \cite{Bronstein_2017, bronstein2021geometricdeeplearninggrids, Gerken:2021sla}. In practice, it turns out that rotation and mirror symmetry are rather difficult to achieve exactly without incurring significant algorithmic overhead (see, e.g., Ref.~\cite{Aronsson:2023rli}). Apart from global symmetries, sampling gauge theories in particular requires the correct implementation of local gauge symmetries. Models that are compatible with gauge symmetry are termed gauge-equivariant \cite{cohen2019gaugeequivariantconvolutionalnetworks}.
One particular architecture relevant for lattice gauge theories, the lattice gauge-equivariant convolutional network (L-CNN) was put forward a few years ago \cite{Favoni:2020reg, Aronsson:2023rli}  by some of the authors of this work. The L-CNN is flexible enough to represent any linear combination of Wilson loops and can therefore, in principle, approximate any gauge-invariant function \cite{Durhuus1980}. The L-CNN is by no means the only gauge-equivariant architecture in the literature: there are gauge-equivariant formulations adapted to solving the Dirac equation \cite{Lehner:2023bba, Lehner:2023prf, Pfahler:2026fpj}, smearing-based formulations \cite{Nagai:2021bhh}, the gauge covariant transformer \cite{Nagai:2025rok}, and implementations based on spectral properties of the gauge group widely used in equivariant NFs (\textit{spectral flow}) \cite{Boyda:2020hsi}.

Turning to generalization, neural lattice samplers need to be able to generalize to different lattice sizes and couplings to be competitive with traditional MC algorithms. Generalization to arbitrary lattice sizes is achievable in principle with all approaches, but the scaling can vary drastically (cf.~direct sampling with NFs vs.~SNFs).
Out-of-distribution sampling at arbitrary coupling constants has, so far, remained rather elusive with a few exceptions \cite{Zhu:2025pmw, Kapust:2025ida, Aarts:2026zzr,Alharazin:2026lcb}. Finally, there should be some guarantee of exactness, i.e., that a neural lattice sampler produces the correct target distribution, like any reasonable MC method. In many approaches, this is indeed the case: samples generated using NFs can be efficiently reweighted to the target density because the Jacobian determinant of the flow is easy to compute. SNFs are similarly exact by construction. For standard DMs, exactness is generally much more difficult to assess, and the model's bias and accuracy should be estimated by comparison with known data. However, once trained, DMs can also be formulated using a deterministic reverse ODE, which enables the same kind of reweighting correction as in CNFs. However, it is generally known that the deterministic formulation of DMs is less accurate than the SDE-based formulation \cite{xu2023restart, deveney2023closingodesdegapscorebased}. 

In this paper, a companion to \cite{Aarts:2026zzr}, we continue the construction of gauge-equivariant score-based diffusion models for SU($N_c$) and U($N_c$) lattice gauge theories, using diffusion-time-dependent L-CNNs as score models.
Ref.~\cite{Aarts:2026zzr} contained first results for SU(2) and U(2) in two spacetime dimensions; here we extend this work by providing the appropriate analytical framework and presenting more numerical results in U(2).

The paper is structured as follows: in Section \ref{sec:ltg_sde_rsde}, we review diffusion processes on non-abelian gauge groups, starting from a general SDE, and derive the Fokker-Planck equation and the associated reverse SDE. We also discuss analytic solutions for the two-dimensional case, which allows us to benchmark the trained models against a known ground truth. In Section \ref{sec:learning}, we discuss how to train diffusion models for sampling lattice gauge theories based on score matching. We use an L-CNN to model the gauge-equivariant score function, which is equivariant under both translations and gauge transformations. A review of the L-CNN and the details of our diffusion model architecture are discussed in Section \ref{sec:lcnn}. MAALA enables our model to sample at arbitrary couplings and is described in Section \ref{sec:maala}. 
Section \ref{sec:numerical_results} finally contains extensive numerical experiments for two-dimensional U(2) lattice gauge theory.
Some technical aspects of the derivations are contained in the Appendices.

\section{Lattice gauge theory and diffusion on non-abelian Lie groups} \label{sec:ltg_sde_rsde}

We begin our discussion with some preliminaries regarding the lattice approximation of non-abelian Yang-Mills theory.
The lattice approximation is based on discretizing continuous $D$-dimensional space-time as a (hyper)cubic lattice $\Lambda $ consisting of lattice sites $x \in \Lambda$ separated by the lattice spacing $a$.
For simplicity, we use $a =1$, expressing everything in lattice units, and assume a periodic lattice of size $L^D$ where $L$ is the number of lattice sites along one lattice axis. Upon discretization, the Euclidean Yang-Mills action is replaced with the Wilson action
\begin{align} \label{eq:wilson_action}
    S_E(\mathbf{U}) = \frac{\beta}{N_c} \sum_{x, \mu < \nu} \mathrm{Re} \, \mathrm{Tr} \,  [\mathds{1} - P_{x, \mu\nu}],
\end{align}
where the inverse coupling $\beta = 2 N_c / g_\mathrm{YM}^2$ is related to the Yang-Mills coupling $g_\mathrm{YM}$ and the number of colors $N_c$. The sum is taken over the lattice sites $x \in \Lambda$ and the lattice axes $0 \leq \mu, \nu < D$. The plaquette $P_{x,\mu\nu}$ is a $1 \times 1$ Wilson loop in the $(\mu\nu)$-plane given by
\begin{align}
    P_{x,\mu\nu} = U_{x, \mu} U_{x+\hat\mu, \nu} U^\dagger_{x+\hat\nu, \mu}  U^\dagger_{x, \nu},
\end{align}
where the gauge link $U_{x,\mu}$ connects site $x$ to $x+\hat \mu$. Here, $\hat \mu$ denotes the unit vector along the $\mu$-axis. We also sometimes denote the plaquette without its indices by $U_P$.
Gauge links and plaquettes are elements of the gauge group $G$ which we take to be either SU$(N_c)$ or U$(N_c)$. A (gauge) link configuration $\mathbf U$ is the set of all gauge links on the lattice.

The lattice approximation retains exact gauge invariance. Given a lattice gauge transformation $\Omega_x \in G$ consisting of arbitrary group elements at each lattice site $x$, we define the gauge transformation of the gauge link as
\begin{align} \label{eq:gauge_tr}
    U'_{x,\mu} = \Omega_x U_{x,\mu} \Omega^\dagger_{x+\hat \mu}.
\end{align}
From this, it follows that the plaquettes transform as
\begin{align}
    P'_{x,\mu\nu} = \Omega_x P_{x,\mu\nu} \Omega^\dagger_{x}.
\end{align}
Consequently, the Wilson action, consisting of only traced plaquettes, is gauge invariant,
\begin{align}
    S_E(\mathbf U') = S_E(\mathbf U).
\end{align}

Expectation values of gauge-invariant observables $O(\mathbf U)$ are defined as the weighted integral over all realizations of the link configurations,
\begin{align}
    \langle O \rangle  =  \frac{1}{\mathcal{Z}}  \int  \mathcal{D}\mathbf{U} \,O(\mathbf{U}) \, e^{-S_E(\mathbf{U})},
\end{align}
where the measure $\mathcal{D}\mathbf{U}$ is the product of the individual Haar measures of each link
\begin{align}
    \mathcal{D}\mathbf{U} = \prod_{x, \mu} \DD U_{x,\mu}.
\end{align}
The partition function $\mathcal{Z}$ guarantees normalization via
\begin{align}
    \mathcal{Z}  = \int \mathcal{D}\mathbf{U} \, e^{-S_E(\mathbf{U})}.
\end{align}
We may write the expectation value as
\begin{align} \label{eq:ev_t0}
    \langle O \rangle  =   \int  \mathcal{D}\mathbf{U} \,O(\mathbf{U}) \, p_0(\mathbf U), \quad p_0(\mathbf U)=\frac{e^{-S_E(\mathbf U)}}{\mathcal{Z}},
\end{align}
where $p_0(\mathbf U)$ is the probability density of the link configurations. In practice, these integrals are computed using Monte Carlo integration
\begin{align}
        \langle O \rangle  \approx   \frac{1}{N_{\rm conf}} \sum_{i=1}^{N_{\rm conf}} O(\mathbf U_i), \qquad \mathbf{U}_i \sim p_0(\mathbf U).
\end{align}
Here, the set of configurations $\{ \mathbf U_1, \mathbf U_2, \dots, \mathbf U_{N_{\rm conf}}  \}$ is distributed according to the probability density $p_0$.

To recover the continuum theory, that is, to make predictions for the original Yang-Mills theory that was discretized on the lattice, we need to send the lattice spacing to zero, $a \rightarrow 0$. Since $a$ is not a parameter that is controlled directly, but set implicitly by $\beta$, the continuum limit is taken by $L \rightarrow \infty$ and $\beta \rightarrow \infty$, such that e.g.,~the physical volume $(La)^D$ is fixed.
The opposite limit, $\beta\to 0$, is the (unphysical) strong-coupling limit, in which the partition function reduces to an integral over the Haar measure, without a Boltzmann weight. 

\subsection{Diffusion process}

We are interested in diffusion processes acting on gauge link configurations and, in particular, how the probability density $p_0(\mathbf U)$ changes under diffusion. We define the group-preserving diffusion process via the Stratonovich stochastic differential equation (SDE)
\begin{align} \label{eq:forward_gauge}
    \dd U^{(t)}_{x,\mu} = i  \left[ \hat K_{x,\mu}(\mathbf U^{(t)}) \, \dd t + g(t) \, \dd \hat W^{(t)}_{x,\mu}\right] \! \circ  U^{(t)}_{x,\mu}.
\end{align}
Here, $t \in [0, T]$ with $T > 0$ is the diffusion time,
$\hat K_{x,\mu}(\mathbf U) \in \mathfrak g$ is the Lie-algebra-valued drift term and $g(t) > 0$ is a time-dependent diffusion coefficient. 
The time interval can be unbounded ($T\to \infty$), but for the application to DMs, we consider a finite extent.
Like the drift term, the stochastic increment $\dd \hat W^{(t)}_{x,\mu}$ is algebra-valued and can be written in terms of group generators \mbox{$T^a \in \mathbb{C}^{N_c \times N_c}$},
\begin{align}
    \dd \hat W^{(t)}_{x,\mu} = \sum^{D_A}_{a=1} T^a  \dd  W^{(t), a}_{x,\mu}, 
\end{align}
where the sum is taken over the color index $1 \leq a \leq D_A$ with $D_A = N_c^2 - 1$ for SU$(N_c)$ and $D_A = N_c^2$ for U$(N_c)$. The generators are Hermitian and normalized as
\begin{align}
    \mathrm{Tr}[T^a T^b] = \frac{1}{2} \delta^{ab}.
\end{align}
The $\dd  W^{(t), a}_{x,\mu}$ are the stochastic increments of independent Wiener processes. The use of the Stratonovich stochastic integral, denoted by the symbol ``$\circ$'' in Eq.~\eqref{eq:forward_gauge}, will simplify certain analytical calculations. To see why, we first write Eq.~\eqref{eq:forward_gauge} more compactly as
\begin{align}
    \dd U^{(t)}_{x,\mu} = i \dd \hat \Theta^{(t)}_{x,\mu} \circ U^{(t)}_{x,\mu},
\end{align}
with
\begin{align}
    \dd \hat \Theta^{(t)}_{x,\mu} &= \hat K_{x,\mu}(\mathbf U^{(t)}) \, \dd t + g(t) \, \dd \hat W^{(t)}_{x,\mu} \nonumber \\
    &= \sum_a T^a \dd  \Theta^{(t), a}_{x,\mu}.
\end{align}
Given a scalar, gauge-invariant function $f(\mathbf U)$, it holds that
\begin{align}
    \dd f^{(t)} = \sum_{a,x,\mu} D^a_{x,\mu} f(\mathbf U^{(t)}) \circ \dd \Theta^{(t), a}_{x,\mu},
\end{align}
where $D^a_{x,\mu}$ is the group derivative (see Appendix \ref{app:group_derivative}) as defined by
\begin{align}
    D^a_{x,\mu} f(\mathbf U) = \frac{\dd }{\dd \epsilon} f(\tilde{\mathbf{U}}) \rvert_{\epsilon = 0},
\end{align}
where $\tilde{\mathbf{U}}$ is the same as $\mathbf U$ except that the link $U_{x,\mu} $ is replaced by the perturbed link $e^{i \varepsilon T^a} U_{x,\mu}$ with $\epsilon > 0$. 
We note that, unlike the Itô integral, the Stratonovich integral preserves the chain rule in the usual sense, see Eqs.~\eqref{eq:app_strat_chain_rule} and \eqref{eq:strat_sde_nonabelian_chain_rule}.

Two specific examples of diffusion processes are well known. In the first case one chooses a drift equal to minus the derivative of a gauge-invariant action $S(\mathbf U)$, e.g., the Wilson action \eqref{eq:wilson_action}, and a constant diffusion coefficient,
\begin{align} \label{eq:sq_drift}
    \hat K_{x,\mu}(\mathbf U) = - T^a D^a_{x,\mu} S(\mathbf U), \qquad g(t) = \sqrt{2}.
\end{align}
This choice leads to the SDE of stochastic quantization \cite{Parisi:1980ys,Damgaard:1987rr}. Starting from an arbitrary initial distribution, in the long-time limit ($T \rightarrow \infty$) the stochastic process approaches a stationary distribution and the gauge links are distributed according to the Boltzmann weight, $\propto \exp(-S(\mathbf U))$.

In the second case, no drift is included at all, $\hat K_{x,\mu}=0$. Starting again from arbitrary initial conditions, the gauge links are randomized and uniformly distributed over the Haar measure. Hence these two processes converge to distributions which are maximally distinct. Below we identify these with the target distribution, $\propto \exp(-S(\mathbf U))$, and the prior (uniform) distribution.

\begin{figure}
    \centering
    \includegraphics[width=0.95\linewidth]{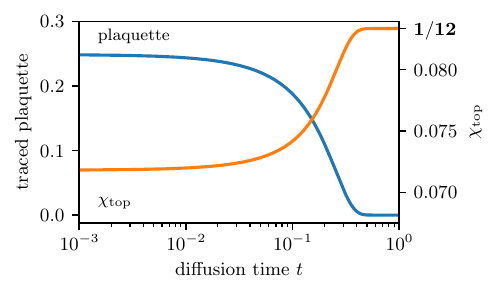}
    \caption{Traced plaquette and topological susceptibility along the diffusion process on a lattice with $\beta=2, L=16$, using the analytical solution for the observables. Here we have chosen no drift and a diffusion coefficient $g(t)=25^{t/T}$ with $T=1$, which leads to $p_t(\mathbf U)$ converging to the uniform distribution (Haar measure). In this limit, the plaquette vanishes and $\chi_{\rm top}=1/12$. }
    \label{fig:forward_diffusion}
\end{figure}

The drift term in Eq.~\eqref{eq:sq_drift} is gauge covariant, i.e., under Eq.~\eqref{eq:gauge_tr}, the drift changes according to
\begin{align} \label{eq:drift_covariance}
    \hat K_{x,\mu}(\mathbf U') = \Omega_x \hat K_{x,\mu}(\mathbf U) \Omega^\dagger_x. 
\end{align}
If the initial conditions at $t=0$ are chosen from a gauge-invariant distribution $p_{\rm in}(\mathbf U') = p_{\rm in}(\mathbf U)$, for example based on the distribution in Eq.~\eqref{eq:ev_t0}, then, together with Eq.~\eqref{eq:drift_covariance}, the stochastic process generates a gauge-invariant probability distribution $p_t(\mathbf U)$ along diffusion time $t > 0$. This holds for any drift and diffusion term, as long as the drift term is covariant.
Observables can then be computed along the stochastic process and depend on the diffusion time. This is illustrated in Fig.~\ref{fig:forward_diffusion} for the traced plaquette and the topological susceptibility (defined below) in the case of U(2) gauge theory on a $L^2=16^2$ lattice at $\beta=2$. Note that $L$ is large enough such that finite-size effects can safely be ignored.  Here the initial distribution is determined by the Wilson action, see Eq.~\eqref{eq:wilson_action}, and no drift is applied. 
The final distribution at $T=1$ is close to the uniform distribution, for which the plaquette vanishes and $\chi_{\rm top}=1/12$.
The computation is performed analytically; details can be found in Appendix \ref{app:analytical_solution}. 
Below we come back to this and similar figures in great depth.

\subsection{Fokker-Planck equation}

Our goal is to derive the Fokker-Planck equation (FPE) associated with the diffusion process in Eq.~\eqref{eq:forward_gauge}. To simplify the discussion, we consider a diffusion process of a single group element $U$ given by
\begin{align} \label{eq:strat_sde_single}
    \dd U^{(t)} &= i  T^a [K^{a}(U) \dd t + g(t) \dd W^{(t), a}] \circ U^{(t)} \nonumber \\
    &= i T^a \dd \Theta^{(t), a} \circ U^{(t)},
\end{align}
with initial conditions $U^{(0)}$ distributed according to some probability density $p_{\rm in}(U)$. The sum over the color index $a$ is implied.
Using the Stratonovich chain rule, the stochastic increment of a scalar function $f$ is given by
\begin{align}
    \dd f^{(t)} &=  D^a f(U^{(t)}) \circ \dd \Theta^{(t), a} \nonumber \\
    &=  (D^a f ) K^{a} \dd t + g(t) D^a f \circ \dd W^{(t), a},
\end{align}
where we have separated the deterministic (left) and the stochastic (right) terms in the last line.

To derive the FPE, we consider a discretization of diffusion time $t$ into finite steps $\epsilon$. We introduce the finite increment
\begin{align}
    \dd f^{(t)} \approx \delta f^{(t)} = f^{(t+\epsilon)} - f^{(t)}.
\end{align}
The deterministic term is approximated in the usual way
\begin{align}
     (D^a f ) K^{a} \, \dd t \approx \left[ D^a f(U^{(t)})\right] K^a(U^{(t)}) \, \epsilon,
\end{align}
whereas the Stratonovich increment is discretized using the mid-point rule
\begin{align} \label{eq:f_strat_term}
     &g(t) D^a f \circ \dd W^{(t), a} \approx \nonumber \\
     &\quad \frac{1}{2} \left[
     g(t+\epsilon) D^a f(U^{(t+\epsilon)}) + g(t) D^a f(U^{(t)}) 
     \right] \nonumber \\
     &\quad \times(W^{(t+\epsilon), a} - W^{(t), a}).
\end{align}
The increment of the Wiener process can be written as
\begin{align}
    W^{(t+\epsilon), a} - W^{(t), a} = \sqrt{\epsilon} \eta^a,
\end{align}
with random variables $\eta^a$ which follow a normal distribution with zero mean and unit variance, i.e.,
\begin{align}
    \langle \eta^a \rangle_\eta = 0, \quad \langle \eta^a \eta^b \rangle_\eta = \delta^{ab}.
\end{align}
Here, $ \langle \dots \rangle_\eta$ denotes the noise average.

The finite increment of Eq.~\eqref{eq:strat_sde_single} can be written as
\begin{align}
    \delta U^{(t)} &= i T^a [ \sqrt{\epsilon} g(t) \eta^a \frac{U^{(t+\epsilon)} + U^{(t)}}{2} 
    \nonumber \\
    & \qquad +  \epsilon K^a(U)U^{(t)}] + \mathcal{O}(\epsilon^{3/2}) \nonumber \\
    &= i T^a \sqrt{\epsilon} g(t) \eta^a U^{(t)} + \mathcal{O}(\epsilon).
\end{align}
The leading order $\sqrt \epsilon$ term of $\delta U^{(t)}$ is then used to expand the Stratonovich term. For this, we use Eq.~\eqref{eq:exp_group_der_trick} and find
\begin{align}
    g(t+\epsilon) D^a f(U^{(t+\epsilon)}) = g(t) [D^a f(U^{(t)}) & \nonumber \\
     + \sqrt{\epsilon}g(t) \eta^b D^b D^a f(U^{(t)})] + \mathcal{O}(\epsilon), &
\end{align}
and thus
\begin{align} \label{eq:f_strat_term2}
     g(t) D^a f \circ \dd W^{(t), a} = \sqrt \epsilon g(t) (D^a f) \, \eta^a  &\nonumber \\
      + \frac{1}{2} g^2(t) \epsilon \eta^a \eta^b D^a D^b f + \mathcal{O}(\epsilon^{3/2}). &
\end{align}
Putting everything together yields
\begin{align}
    \delta f^{(t)} &=  \left[K^a \, D^a f + \frac{1}{2} g^2(t) \eta^a \eta^b D^a D^b f \right] \epsilon  \nonumber \\
    &\quad + g(t) (D^a f) \, \eta^a \sqrt{\epsilon} + \mathcal{O}(\epsilon^{3/2}),
\end{align}
where the drift $K^a$ and the function $f$ (along with its derivatives) are all evaluated at $U^{(t)}$. Performing the noise average removes the terms linear in $\eta$. We find
\begin{align}
    \langle \delta f^{(t)} \rangle_\eta = \langle K^a \, D^a f \! + \! \frac{1}{2} g^2(t) D^a D^a f  \rangle_\eta \, \epsilon \! +\!  \mathcal{O}(\epsilon^{3/2}).
\end{align}
Taking the $\epsilon \rightarrow 0$ limit, we obtain the time derivative of the expectation value of $f$,
\begin{align} \label{eq:dfdt_eta}
    \frac{d}{dt} \langle f \rangle_\eta = \langle  K^a \, D^a f \rangle_\eta + \frac{1}{2} g^2(t) \langle D^a D^a f  \rangle_\eta.
\end{align}
The average over the noise can be replaced by an integration over a time-dependent probability distribution $p_t(U)$,
\begin{align}
    \langle f \rangle_\eta = \langle f \rangle_t \equiv \int \DD U \, f(U) \, p_t(U), 
\end{align}
where $\DD U$ is the Haar measure. Replacing all three terms with their expectation values w.r.t.~$p_t$ and using integration by parts (see Eq.~\eqref{eq:group_der_int_parts}) yields
\begin{align}
    \frac{d}{dt} \langle f \rangle_\eta &= \int \DD U \, f(U) \frac{d}{dt} p_t(U), \\
    \langle K^a D^a f  \rangle_\eta &= -\int \DD U \, f(U) D^a (K^a(U) p_t(U)), \\
    \langle D^a D^a f  \rangle_\eta &= \int \DD U \, f(U) D^a D^a p_t(U).
\end{align}
Inserting this into 
Eq.~\eqref{eq:dfdt_eta} and using the fact that the equation must hold for any function $f$, we arrive at the FPE
\begin{align} \label{eq:fpe_single}
    \frac{d}{dt} p_t(U) &= -D^a[K^a(U) p_t(U)] + \frac{1}{2} g^2(t) \hat \Delta \,  p_t(U), 
\end{align}
where $\hat \Delta = D^a D^a$ is the Laplace-Beltrami operator. 

It is trivial, but cumbersome, to repeat the derivation for the process on the lattice, Eq.~\eqref{eq:forward_gauge}. The FPE reads
\begin{align} \label{eq:fpe_lattice}
        \frac{d}{dt} p_t(\mathbf U) &=  -\sum_{a,x,\mu} D^a_{x,\mu}[ K^a_{x,\mu}( \mathbf U) p_t(\mathbf U) ] 
        + \frac{1}{2} g^2(t) \hat \Delta \, p_t(\mathbf U), 
\end{align}
with the Laplace-Beltrami operator
\begin{align}
\label{eq:LB}
    \hat \Delta \,  p_t(\mathbf U) = \sum_{a, x,\mu} D^a_{x,\mu} D^a_{x,\mu} \, p_t(\mathbf U).
\end{align}

A forward process with a gauge-covariant drift term, see Eq.~\eqref{eq:drift_covariance}, leads to a gauge-invariant FPE. Both the divergence term, $D^a (K^a p_t)$, and the Laplace-Beltrami term are gauge invariant. Consequently, the solutions of the FPE inherit the gauge invariance property. 

In the case of stochastic quantization, i.e.,~Eq.~\eqref{eq:sq_drift}, the FPE becomes
\begin{align} \label{eq:fpe_sq_lattice}
       \! \frac{d}{dt} p_t(\mathbf U) &= \sum_{a,x,\mu} \! D^a_{x,\mu}[ D^a_{x,\mu} S( \mathbf U) p_t(\mathbf U) ] + \hat \Delta \,  p_t(\mathbf U) \nonumber \\
        &= \sum_{a,x,\mu} \! D^a_{x,\mu}[ (D^a_{x,\mu} S( \mathbf U))  +  D^a_{x,\mu}] p_t(\mathbf U).
\end{align}
If the system thermalizes (at $t \rightarrow \infty$), the time derivative on the left-hand side vanishes. 
In this stationary limit, it is easy to see from the second line of Eq.~\eqref{eq:fpe_sq_lattice} that the stationary FPE admits the solution
\begin{align}
    p_\infty(\mathbf U) = \frac{1}{\mathcal Z} e^{-S(\mathbf U)},
\end{align}
i.e., the target ensemble is obtained. 

\subsection{Analytical analysis of the forward process}\label{sec: analytical_diffusion}

We consider the FPE (\ref{eq:fpe_lattice}) without a drift term, $K^a_{x,\mu}(\mathbf U) \equiv 0$. The evolution of the distribution $p_t(\mathbf U)$ is then determined by the heat equation
\begin{equation}
        \frac{d}{dt} p_t(\mathbf U) = \frac{1}{2} g^2(t) \hat \Delta \,  p_t(\mathbf U), 
	\label{eq:heat-equation}
\end{equation}
where $\hat \Delta$ is the Laplace-Beltrami operator (\ref{eq:LB}), 
 written here as a sum of Laplacians acting on links $\ell=(x,\mu)$,
\be
\label{eq:Delta_l}
 \hat \Delta = \sum_\ell \hat \Delta_\ell, \qquad 
  \hat \Delta_\ell = \sum_{a} D^a_{\ell} D^a_{\ell}.
\ee
As initial condition we take the target distribution, $p_{0}(\mathbf U) = e^{-S_E(\mathbf U)}/{\mathcal Z}$, with $S_E$ the Wilson action (\ref{eq:wilson_action}) and gauge group U($N_c$) for definiteness.

In the following, we rescale the time variable $t$ according to the noise profile,
\be
\label{eq:time}
\tau(t) = \int_0^t du\, g^2(u),
\ee
which absorbs the factor of $g^2(t)$ in Eq.~(\ref{eq:heat-equation}).

Since the Laplacian $\hat\Delta$ commutes with the generators of the algebra $\mathfrak{u}(N_c)$, Schur's lemma implies that it acts as a constant on each irreducible representation of U($N_c$).
Indeed, the action of the Laplacian on character functions of representation $r$, $\chi_r(\textbf{U}) = \mathrm{Tr}_r(\textbf{U})$, is the action of the quadratic Casimir operator,
\begin{equation}
    \hat \Delta_\ell \chi_r(\textbf{U}) = -C_2(r)\chi_r(\textbf{U}),
\end{equation}
and hence
\begin{equation}\label{eq: semigroup_on_link}
   e^{\frac{\tau}{2} \hat \Delta_\ell} \chi_r(\textbf{U}) = e^{-\frac{\tau}{2} C_2(r)}\chi_r(\textbf{U}).
\end{equation}
The character functions $\chi_r(\textbf{U})$ form a basis over representations. The Peter-Weyl theorem helps to construct a general solution that satisfies Eq.~(\ref{eq:heat-equation}) with a delta distribution as the initial condition, $\rho_{\tau=0}(\textbf{U}) = \delta(\textbf{U})$, 
\begin{equation}
	\rho_\tau(\mathbf U) = \sum_{r} d_r e^{-\frac{\tau}{2}C_2(r)} \chi_r(\mathbf U),
\end{equation}
where \(d_r \) is the dimension of the representation \(r\).

This heat kernel describes the action of the semigroup \(e^{\frac{1}{2}\tau \hat\Delta } \) on the probability distribution \(p_0(\mathbf U)\) through convolution,
\begin{equation}
	p_\tau (\mathbf U) \equiv e^{\frac{\tau}{2}  \hat \Delta_\ell }p_0(\mathbf U) = \int \dd\mathbf V \rho_\tau(\mathbf U \mathbf V^{-1}) p_0(\mathbf V).
\end{equation}
To evaluate the effect of diffusion, we note that the action of the Laplacian on the lattice induces a Brownian motion independently on each link. Since each plaquette $U_P$ is composed of four links $\ell$ and, on a two-dimensional lattice with periodic boundary conditions, each link is shared by two neighboring plaquettes, with opposite orientation, the diffusion process will entangle the plaquettes; we verify this explicitly below. 

Using the fact that the Laplacian is self-adjoint under the Haar measure, the action of the Laplacian can be moved from the distribution to observables, according to
\begin{align}
\bra O\ket_\tau  &= \int {\cal D}U \, p_\tau(\mathbf U) O(\mathbf U) 
= \int {\cal D}U \, \left( e^{\frac{\tau}{2}\hat \Delta}p_0(\mathbf U) \right) O(\mathbf U) \nn\\
&= \int {\cal D}U \, p_0(\mathbf U) \left( e^{\frac{\tau}{2}\hat \Delta} O(\mathbf U) \right).
\end{align}
We now consider as observables non-self-intersecting loops $\Gamma$ of perimeter $\mathcal P(\Gamma)$,
\be
W_\Gamma^{(r)} = \frac{1}{d_r} \mathrm{Tr}_r U_\Gamma.
\ee
Obvious examples are rectangular Wilson loops of size $R\times T$ in representation $r$ with perimeter $\mathcal P(\Gamma)=2(R+T)$.
Since each link only appears once, we find under forward diffusion a simple exponential time dependence with the rate determined by the perimeter of the closed loop,
\begin{equation}
 e^{\frac{\tau}{2}  \hat \Delta} W_\Gamma^{(r)} = e^{-\frac{\tau}{2}  \mathcal P(\Gamma)C_2(r)} W_\Gamma^{(r)},
\end{equation}
or for its expectation value
\begin{equation}
  \bra W_\Gamma^{(r)} \ket_\tau  = e^{-\frac{\tau}{2} \mathcal P(\Gamma)C_2(r)} \bra W_\Gamma^{(r)} \ket_0.
\end{equation}
This time dependence is independent of the dimensionality and volume of the lattice (as long as the loop is edge-simple, i.e., each link only appears once) and also of the gauge coupling. 

An immediate consequence is that the Creutz ratio \cite{Creutz:1980zw} is exactly preserved under forward diffusion, 
\begin{align}
\chi^{(r)}_\tau(R,T) &= -\log \frac{
\bra W^{(r)}_{R,T}\ket_\tau 
\bra W^{(r)}_{R+1,T+1}\ket_\tau }{
\bra W^{(r)}_{R+1,T}\ket_\tau 
\bra W^{(r)}_{R,T+1}\ket_\tau }
\nn\\
&= \chi^{(r)}_0(R,T),
\label{eq:Creutz}
\end{align}
since the time-dependent exponentials cancel. 
Hence the string tension is not affected by the diffusion process, i.e., this information is not removed in the noising process. This phenomenon is very similar to the conservation of higher-order cumulants on non-compact flat manifolds in the variance-expanding scheme \cite{Aarts:2024rsl}. 
We come back to the behaviour of Creutz ratios in the numerical implementation of the (de)noising processes below.

\begin{figure}
    \centering
    \includegraphics[width=0.99\linewidth]{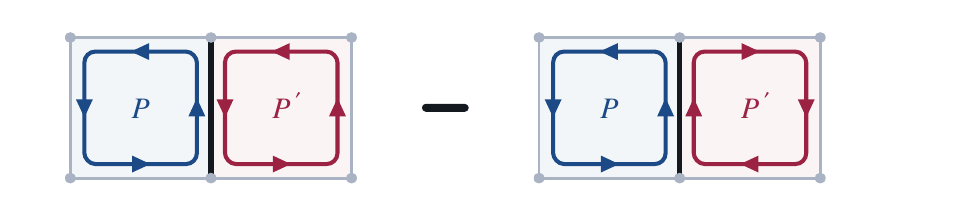}
    \caption{Products $O_\pm$ of two traced plaquettes $U_P, U_P'$ sharing one link, with different orientations, see Eq.~(\ref{eq:plaq_correlators}). The difference, $\bra O_+\ket -\bra O_-\ket$, vanishes for the target Wilson action in two dimensions, but becomes non-zero during the diffusion process, indicating the non-locality of the time-dependent probability distribution. Figure created with Claude Opus 5.5.
    }
    \label{fig:OpOm}
\end{figure}

While the forward evolution of these edge-simple loops is particularly easy, this is not the case for the distribution $p_\tau(\mathbf U)$ or its log derivative (which is the reason the score needs to be learned, after all). This can be demonstrated by considering as observables the products of two plaquettes, $U_P, U_P'$, which have one link in common. We consider the observables 
\begin{align}
    \begin{split}
    O_+ &= \frac{1}{N_c^2} \Re\, \tr{U_P} \tr{U_{P'}}, \\
    O_- &= \frac{1}{N_c^2} \Re\, \tr{U_P} \tr{{U_{P'}}^\dagger},\\
    O_P &= \frac{1}{N_c} \Re\, \tr{U_P}, 
    \end{split} \label{eq:plaq_correlators}
\end{align}
where $U_P$ and $U_P'$ share one link, see Fig.~\ref{fig:OpOm}.
For the Wilson plaquette action, i.e., at $\tau=0$, in the large volume limit, the expectation values of these are, in two dimensions, equal to the square of the plaquette expectation value, 
\begin{align}
    \langle O_+ \rangle_0 = \langle O_- \rangle_0 = \frac{1}{N_c^2}\langle \tr{U_P} \rangle_0^2 
    =\bra O_P\ket_0^2,
\end{align}
and hence the difference vanishes. Under forward diffusion the action of the Laplace-Beltrami operator on the products of links results in a cross term, which is different for $O_\pm$, and
\begin{align}
    \langle O_+ \rangle_\tau - \langle O_- \rangle_\tau \neq 0 \quad\mbox{when}\quad \tau>0.
\end{align}
Note that for SU(2) the traced plaquette is real and $O_+=O_-$; instead one can consider
\begin{align}
\label{eq:Opm2}
    \langle O_\pm \rangle_\tau - \langle O_P\rangle_\tau^2 \neq 0 \quad\mbox{when}\quad \tau>0.
\end{align}   
The difference from $0$ implies that the time-dependent probability distribution no longer factorises in plaquettes. 
This observation is shown numerically in Section~\ref{sec:time-dep-action}.

\subsection{Reverse diffusion process}

To determine the reverse diffusion process, we first consider the time-reversal of the FPE \eqref{eq:fpe_lattice}. Assuming that $p_t(\mathbf U)$ solves the FPE, we can define the time-reversed probability density
\begin{align}
    \tilde p_{\tilde t}(\mathbf U) =  p_{T - \tilde t}(\mathbf U), \qquad \tilde t = T - t \in [0, T].
\end{align} 
The time derivative of $\tilde p_{\tilde t}$ leads to an ill-posed FPE with a wrong sign in front of the diffusion term,
\begin{align}
    \frac{d}{d\tilde t} \tilde p_{\tilde t} = \sum_{a,x,\mu} D^a_{x,\mu}[ K^a_{x,\mu} \, \tilde p_{\tilde t} ] - \frac{1}{2} \tilde g^2(\tilde t) \hat \Delta \,  \tilde p_{\tilde t},
\end{align}
with $\tilde{g}(\tilde t) = g(T -  \tilde t)$. By adding and subtracting $\tilde g^2(\tilde t) \hat \Delta \,  \tilde p_{\tilde t}$, the time-reversed FPE can be written as
\begin{align}
    \frac{d}{d\tilde t} \tilde p_{\tilde t} &= -\sum_{a,x,\mu} D^a_{x,\mu} \left[ \left( \tilde g^2(\tilde t) D^a_{x,\mu} \ln \tilde{p}_{\tilde{t}} - K^a_{x,\mu} \right) \! \tilde p_{\tilde t} \, \right] \nonumber \\
    \label{eq:fpebw}
    &\qquad+ \frac{1}{2} \tilde g^2(\tilde t) \hat \Delta \,  \tilde p_{\tilde t}.
\end{align}
Here, we have used that
\begin{align}
    \tilde g^2(\tilde t) \hat \Delta \,  \tilde p_{\tilde t} = \sum_{a,x,\mu} D^a_{x,\mu} \left[ \left( \tilde g^2(\tilde t) D^a_{x,\mu} \ln \tilde p_{\tilde t} \right) \tilde p_{\tilde t} \right].
\end{align}
This FPE, with the re-interpreted drift term on the first line of Eq.~(\ref{eq:fpebw}), is now well-posed. 

The initial conditions for the reverse FPE at $\tilde t = 0$ must be chosen in such a way that they are compatible with the forward solution of the FPE \eqref{eq:fpe_lattice} at time $t=T$, i.e., 
$\;\tilde p_{\tilde 0}(\mathbf U) \sim p_T(\mathbf U)$. A precise agreement is not required, since the presence of the drift term in Eq.~(\ref{eq:fpebw}) ensures memory of the initial conditions is erased.

The associated \textit{reverse diffusion process} \cite{anderson1982reverse} is then 
\begin{align}\label{eq:reverse_gauge}
     & \dd U^{(\tilde t)}_{x,\mu} =  i \sum_a T^a \big[ \big( - K^a_{x,\mu}(\mathbf U^{(\tilde t)}) 
      \\
       \nonumber 
     &\quad  + \tilde g^2(\tilde t) D^a_{x,\mu} \ln \tilde p_{\tilde t} (\mathbf U^{(\tilde t)}) \big) \dd \tilde t 
    + \tilde g(\tilde t) \dd  \tilde W^{(\tilde t), a}_{x,\mu} \big] \circ U^{(\tilde t)}_{x,\mu},
\end{align}
with initial conditions $\mathbf U^{(\tilde 0)} \sim \tilde p_{\tilde 0}(\mathbf U)$. Here, $\dd \tilde W^a_{x,\mu}$ are the increments of reversed Wiener processes.

\section{Learning the diffusion process} \label{sec:learning}

In contrast to the forward diffusion process, the reverse diffusion equation \eqref{eq:reverse_gauge} requires knowledge of the derivative of the logarithm of the time-dependent probability density $D^a_{x,\mu} \ln p_t(\mathbf U)$. If this derivative were known, solving the reverse SDE would enable sampling the target density $p_0(\mathbf U)$. Given that exactly solving for $ p_t(\mathbf U)$ through the Fokker-Planck equation is, in general, intractable, diffusion models require machine learning techniques to find approximations to this derivative, the so-called score function $s_{\theta;x,\mu}^a(\mathbf{U}, t)$. The machine-learning task is to find a suitable score function, typically parameterized by a neural network with parameters $\theta$, such that
\begin{align} 
    s_{\theta;x,\mu}^a(\mathbf{U}, t) \approx  D^a_{x,\mu} \ln p_t(\mathbf U),
\end{align}
for all $t \in [0, T]$ and all link configurations $\mathbf U$. Writing the time-dependent probability as
\begin{align} \label{eq:St}
    p_t(\mathbf U) = \frac{1}{\mathcal Z(t)} \exp(-\mathcal S_t(\mathbf U)),
\end{align}
where $\mathcal Z(t)$ is a normalization constant, we see that the optimal score function is just the negative gradient of a time-dependent action $\mathcal S_t(\mathbf U)$. In particular, at $t=0$, the score function can be identified with the negative gradient of the target action.

It is also apparent that the optimal score function inherits all symmetries of the lattice field theory, provided that the Fokker-Planck evolution conserves these symmetries. Thus, the score function should transform locally under lattice gauge transformations, in the same way the gradient of the action transforms, see e.g.~Eq.~(\ref{eq:drift_covariance}), 
\begin{align} \label{eq:score_gauge_transform}
     \hat s_{\theta;x,\mu}(\mathbf{U}', t) =   
    \Omega_x \, \hat s_{\theta;x,\mu}(\mathbf{U}, t)  \Omega^\dagger_x,
\end{align}
where $\hat s_{\theta;x,\mu} = \sum_a T^a s^a_{\theta;x,\mu}$. Furthermore, the score function needs to be compatible with all other global symmetries, such as translations, discrete rotations, and mirror symmetry. It is therefore essential that the score function is parametrized by a gauge-equivariant neural network. 

\subsection{Loss function}
To find the optimal score function $s_{\theta}$, we use denoising score matching (DSM) \cite{6795935}, adapted to gauge groups. In DSM we seek to minimize the loss function
\begin{align} \label{eq:intractable_loss}
    \mathcal{L} = \int_0^T {\dd t}\,  \lambda(t) \underset{ \substack{ p_t(\mathbf{U})}}
    {\mathbb{E}}\sum_{x,\mu,a} \Big[ & s_{\theta;x,\mu}^a(\mathbf{U}^{(t)}, t) &
    \nonumber \\
     & - D^a_{x,\mu} \ln p_{t}(\mathbf U^{(t)}) \Big]^2,
\end{align}
with the expectation value $\mathbb{E}[\dots]$ taken over the probability density $p_t(\mathbf U)$ at time $t$ and $\lambda(t)$ is a strictly positive, time-dependent function, which will be related to the diffusion coefficient below. 
With Eq.~\eqref{eq:score_gauge_transform}, it is clear that the loss function is a gauge invariant quantity.

However, Eq.~\eqref{eq:intractable_loss} requires $p_t(\mathbf U)$, which is generally unknown. In DSM, the loss function is therefore rewritten in the following way,
\begin{align}
    \label{eq:gauge_loss_function}
    \mathcal{L} = \int_0^T {\dd t}\,   \lambda(t) \!\!\!\!\!
    \underset{ \substack{ p_0(\mathbf{U}^{(0)}) \\ p_{0t}(\mathbf{U}^{(t)}|\mathbf{U}^{(0)}) }}
    {\mathbb{E}} 
       \sum_{x,\mu,a} \Big[  s_{\theta;x,\mu}^a(\mathbf{U}^{(t)}, t)  \quad &
       \nonumber\\
        - D^a_{x,\mu} \! \ln p_{0t}(\mathbf U^{(t)} | \mathbf{U}^{(0)}) \Big]^2 + \mbox{const}, &
\end{align}
with $p_{0t}(\mathbf{U}^{(t)}|\mathbf{U}^{(0)})$ the transition probability, i.e., the probability to reach $\mathbf{U}^{(t)}$ at diffusion time $t$ starting with $\mathbf{U}^{(0)}$ as initial condition. Here, the derivative $D^a_{x,\mu}$ acts on the $\mathbf U^{(t)}$ argument of $\ln p_{0t}$.
The two representations of the loss function differ by a $\theta$-independent constant, which is irrelevant for optimizing the score function. In contrast to Eq.~\eqref{eq:intractable_loss}, the
loss function in Eq.~\eqref{eq:gauge_loss_function} is tractable: the expectation value with respect to $p_0$ is replaced by Monte Carlo sampling $\mathbf{U}^{(0)}$ from a training dataset, and for each $\mathbf{U}^{(0)}$, $\mathbf{U}^{(t)}$ is obtained by sampling the forward process numerically. The remaining complication is determining the derivative of the transition probability analytically. For certain simple forward processes, $D^a_{x,\mu} \ln p_{0t}$ may be determined exactly.

\subsection{Forward process and transition probability}

For our calculations, we require a discretized version of the Stratonovich SDE, Eq.~\eqref{eq:forward_gauge}, which is given by
\begin{align} \label{eq:strat_sde_nonabelian_update}
    U^{(t+\epsilon)}_{x,\mu} = \exp \left( i T^a \left[ \epsilon \, K^a_{x,\mu}(\mathbf{U}^{(t)}) + \sqrt{\epsilon} \, g(t) \, \eta^a_{x,\mu}(t) \right] \right) U^{(t)}_{x,\mu},
\end{align}
where $\eta^a_{x,\mu}(t)$ are independent Gaussian random variables with zero mean and unit variance, and the sum over the color index $a$ is assumed.
Details on the Wiener process and Gaussian noise can be found in Appendix~\ref{app:stratonovich}.
Specifically, we use the variance-expanding SDE scheme, where we set $K^a_{x,\mu}(U^{(t)})=0$ and $g(t)=\xi^{t/T}$ (with $T=1$ and $\xi=25$ in the numerical experiments), such that
\begin{align} \label{eq:strat_sde_nonabelian_update_ve}
    U^{(t+\epsilon)}_{x,\mu} = \exp \left[ i \sqrt{\epsilon} \, \xi^{t/T} \, T^a \eta^a_{x,\mu}(t) \right] U^{(t)}_{x,\mu}.
\end{align}
Next, we require the finite-time transition probability of the forward process.
It is instructive to first consider the Abelian case (gauge group $U(1)$) for a single degree of freedom. We consider the process
\begin{align}
    U^{(t+\epsilon)} = \exp \left[ i \sqrt{\epsilon} \, \xi^{t/T}  \eta(t) \right] U^{(t)},
\end{align}
where $|U^{(t)}| = 1$ $\forall t \in [0, T]$.
Because the group elements commute, the update scheme can be iterated  (with $\epsilon \rightarrow 0$) and the process can be written as a single-step update, assuming continuous time, 
\begin{align}
\label{eq:s}
    U^{(t)} = \exp \big[ i \sigma(t) \eta' \big] U^{(0)}, 
    \end{align}
where $\eta'$ is standard Gaussian noise and $\sigma(t)$ is the square root of the rescaled time already introduced in Eq.~(\ref{eq:time}), 
\begin{equation}
\label{eq:sigma2}
 \sigma^2(t) = \int_0^t du\, g^2(u) = \int_0^t du\, \xi^{2u/T} = 
  T\frac{\xi^{2 t/T} - 1}{\ln \xi^2},
\end{equation}
now conventionally denoted with $\sigma^2(t)$.
Using the angle representation, where $\phi^{(t)} = \arg U^{(t)} \in [-\pi, \pi)$, the transition probability is given by the wrapped Gaussian
\begin{align} \label{eq:wrapped_gaussian}
    p_{0t}(\phi^{(t)} | \phi^{(0)}) \!=\! \frac{1}{\sqrt{2\pi} \sigma(t)} \sum_{k \in \mathbb Z} \exp(\!-\frac{[\phi^{(t)} \! - \! \phi^{(0)} \! - \! 2\pi k]^2}{2 \sigma^2(t)}  ).
\end{align}
The sum accounts for all possible $2\pi$-wrappings on the unit circle.

\begin{widetext}
Following the standard trick to derive the DSM loss function, we insert Eq.~\eqref{eq:wrapped_gaussian} into Eq.~\eqref{eq:gauge_loss_function}  and obtain
    \begin{align} \label{eq:wrapped_loss}
            \mathcal{L} &= \int_0^T  {\dd t} \,  \lambda(t) 
            \int_{-\pi}^{\pi} d \phi^{(0)} 
            \int_{-\pi}^{\pi} d \phi^{(t)} \,
        p_0(\phi^{(0)}) \, p_{0t}(\phi^{(t)} | \phi^{(0)}) \, 
        \Big[ 
        s_{\theta}(\phi^{(t)}, t)  - \partial_{\phi^{(t)}} 
        \ln p_{0t}(\phi^{(t)} | \phi^{(0)}) \Big]^2 + \mbox{const}
        \nonumber \\
       &= \int_0^T  {\dd t} \,  \lambda(t)  
       \int_{-\pi}^{\pi} d\phi^{(0)} \, p_0(\phi^{(0)}) \,
       \int_{-\pi}^{\pi} d \phi^{(t)} \,
       \sum_{k \in \mathbb Z}   
       \frac{1}{\sqrt{2\pi} \sigma(t)}
       e^{ -\frac{(\phi^{(t)} - \phi^{(0)} - 2\pi k)^2}{2 \sigma(t)^2}}   
       \left[ 
       s_{\theta}(\phi^{(t)}, t) + 
       \frac{\phi^{(t)} -\phi^{(0)} - 2\pi k }{\sigma^2(t)} 
       \right]^2 + \mbox{const}'.
    \end{align}
In the last line, we have completed the square and absorbed a $\theta$-independent term into the constant. To see this explicitly, we expand the square in the first line of Eq.~\eqref{eq:wrapped_loss} and rewrite the $s^2_\theta$ term as
\begin{align}
    p_{0t}(\phi^{(t)} | \phi^{(0)}) \, s^2_{\theta}(\phi^{(t)}, t) = \frac{1}{\sqrt{2\pi} \sigma(t)} \sum_{k \in \mathbb Z} \exp(\!-\frac{[\phi^{(t)} \! - \! \phi^{(0)} \! - \! 2\pi k]^2}{2 \sigma^2(t)}  ) s^2_{\theta}(\phi^{(t)}, t).
\end{align}
Similarly, for the cross term, we get
\begin{align}
    p_{0t}(\phi^{(t)} | \phi^{(0)}) \, s_{\theta}(\phi^{(t)}, t) \partial_{\phi^{(t)}} 
        \ln p_{0t}(\phi^{(t)} | \phi^{(0)}) &= \, s_{\theta}(\phi^{(t)}, t) \partial_{\phi^{(t)}} 
        p_{0t}(\phi^{(t)} | \phi^{(0)}) \nonumber \\
        &= \frac{1}{\sqrt{2\pi} \sigma(t)} \sum_{k \in \mathbb Z} \exp(\!-\frac{[\phi^{(t)} \! - \! \phi^{(0)} \! - \! 2\pi k]^2}{2 \sigma^2(t)}  ) s_{\theta}(\phi^{(t)}, t) \left[- \frac{\phi^{(t)} \! - \! \phi^{(0)} \! - \! 2\pi k}{\sigma^2(t)}\right].
\end{align}
The term quadratic in $\partial \ln p_{0t}$ is $\theta$-independent and therefore does not contribute to the parameter gradient used during training. 

Assuming that the score function is $2\pi$-periodic, we replace $\phi^{(t)}$ with the new integration variable $\eta = (\phi^{(t)} - \phi^{(0)} - 2\pi k)/\sigma(t)$. Combining the integration range $[-\pi, \pi)$ with the sum over $k$ yields an integral over $\eta$ which covers the real line $\mathbb R$ exactly once. The loss function is then
\begin{align} \label{eq:loss_dsm}
   \mathcal{L}  &= \int_0^T  {\dd t} \,  \lambda(t)  
   \int_{-\pi}^{\pi} \dd \phi^{(0)} \, p_0(\phi^{(0)}) 
   \int_{-\infty}^{\infty} \dd \eta \, 
   \frac{1}{\sqrt{2\pi}} e^{-\eta^2/2}   
    \left[ s_{\theta}(\phi^{(0)} + \sigma(t) \eta, t) 
    + \frac{\eta }{\sigma(t)} 
    \right]^2 
    + \mbox{const}' \nonumber \\
    &= \int_0^T {\dd t} \,  \lambda(t)  
    \underset{ \substack{ p_0(\phi^{(0)}) \\ p_{\eta}(\eta) }}
    {\mathbb{E}} 
    \left[ s_{\theta}(\phi^{(0)} + \sigma(t) \eta, t) 
    + \frac{\eta }{\sigma(t)} \right]^2 + \mbox{const}',
\end{align}
where $\eta$ is drawn from the normal distribution $p_\eta$ with variance 1 and mean 0.
The loss function can be readily computed by first sampling $\phi^{(0)}$ from the training dataset and then sampling random noise $\eta$. Although we used compact degrees of freedom on the unit circle, the loss function in Eq.~\eqref{eq:loss_dsm} is the well-known DSM loss function originally formulated for non-compact variables \cite{6795935}.
\end{widetext}

\begin{figure}
    \centering
    \includegraphics[width=0.95\linewidth]{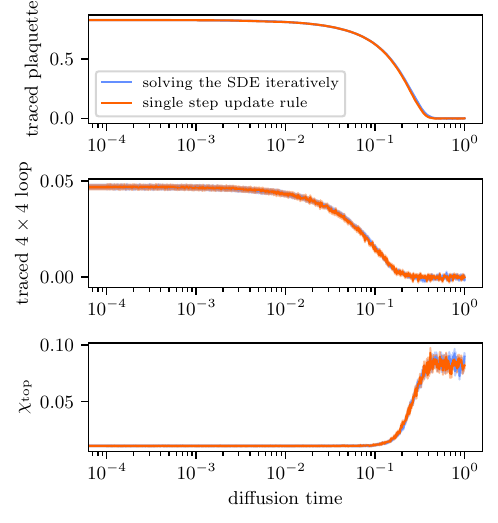}
    \caption{Comparison of simulating the forward diffusion process by solving the forward SDE iteratively and by using the single-step approximation. Shown are three observables (expectation values of the traced plaquette and $4\times 4$ Wilson loop, topological susceptibility) along the forward diffusion process on a $16^2$ lattice, starting from HMC samples generated at $\beta=12$. Good agreement is observed.}
    \label{fig:sde_vs_p0t}
\end{figure}

For the non-abelian process in Eq.~\eqref{eq:strat_sde_nonabelian_update}, the exact single-step update is unknown, but we still use an analogous expression as an approximate way to calculate the evolution of samples under the diffusion process in a single step, namely 
\begin{align} \label{eq:single_step_diffusion_ve}
    {U}^{(t)}_{x,\mu}  \approx \exp \big[ i \sigma(t) \sum_a T^a  \eta^a_{x,\mu}  \big] {U}^{(0)}_{x,\mu}, 
\end{align}
with $\sigma(t)$ given in Eq.~(\ref{eq:sigma2}).
In Fig.~\ref{fig:sde_vs_p0t}, we demonstrate that in terms of the observables we consider in this work (Wilson loops and the topological susceptibility), this approximation does not introduce any noticeable errors, even at large diffusion times.
At very small $\sigma$, i.e, at the start of the forward process,  the Baker-Campbell-Hausdorff corrections are negligible. At very large $\sigma$, the single-step update sufficiently randomizes the initial link $U^{(0)}$ such that the whole group manifold is covered. Deviations are only expected for moderately large $\sigma$, at intermediate times. 
As it provides a significant speedup, we use this approximation for training. 

Similarly, for the DSM loss function, we disregard non-abelian effects and generalize the Abelian result to 
\begin{equation}
    \mathcal{L} = \int_0^T   {\dd t} \, \lambda(t) 
    \underset{ \substack{ p_0(\mathbf{U}^{(0)}) \\ p_{\eta}(\eta) }}
    {\mathbb{E}}
       \sum_{x,\mu,a} \left[ s_{\theta;x,\mu}^a(\mathbf{U}^{(t)}, t) + \frac{\eta_{x,\mu}^a}{\sigma(t)} \right]^2,
\end{equation}
where $\eta^a_\mu$ corresponds to the noise used in Eq.~\eqref{eq:single_step_diffusion_ve}.
The conventional choice of $\lambda(t)$ is to take $\lambda(t)=\sigma^2(t)$, to cancel the $1/\sigma(t)$ in the second term. In this study, we have used $\lambda(t)=\sigma(t)$ instead, and as we will demonstrate below, this has no adverse effect on the training. 
The benefit of using the above loss function is that it can be computed very efficiently, enabling fast training of diffusion models. We do not attempt to quantify the error analytically here, and instead rely on numerical experiments to check whether the approximations are justified.  
In Section \ref{sec:numerical_results}, we see that the numerical results demonstrate, at least for the considered observables, that the errors must be small, within the statistical precision. While Fig.~\ref{fig:sde_vs_p0t} already suggests that the single-step update is sufficiently accurate, we also consider more extensive numerical tests using the reverse SDE. Finally, the MAALA sampler, introduced in Section \ref{sec:maala}, allows for sampling the reverse process even if the score function is only approximate. Thus, by virtue of the Metropolis correction step, systematic errors introduced by the Abelian approximation are removed.

\section{gauge-equivariant neural networks} \label{sec:lcnn}

To maintain gauge invariance, the score function needs to be modeled by a gauge-equivariant neural network. We use a time-dependent Lattice gauge-equivariant Convolutional Neural Network (L-CNN) \cite{Favoni:2020reg}, which naturally fulfills this symmetry, as well as translational symmetry. 

The core of the L-CNN consists of lattice gauge-equivariant bilinear (L-Bilin) and convolutional (L-Conv) layers which take the local gauge symmetry of the lattice into account. The inputs of these layers are the non-locally transforming links $U$ and some locally transforming variables $W$, which we choose to be the plaquettes. 
The L-Bilin layer acts only on the locally transforming variables and is defined as 
\begin{align}
     W'^{l}_{x} = \sum_{m,n} \alpha^{lmn} \,   W^m_{x} \,  W^n_{x}.
\end{align}
Here $m$ and $n$ correspond to the input channels, $1\le m,n \le N_{in}$, and $l$ to the output channel, $1\le l \le N_{out}$. $\alpha^{lmn} \in \mathds{R}$ are the trainable parameters of the layer. The L-Conv layer, which differs from standard convolutional layers by transporting the locally transforming variables within kernel range to the resulting site of the convolution, is given by 
\begin{align}
     W'^{i}_{x} = \sum_{j, k, \mu} \omega^{ij}_{k,\mu} \, \, U_{x, k \hat \mu}  W^j_{x+ k \hat \mu}  U^\dagger_{x, k \hat \mu},
\end{align}
with $1\le i \le N_{out}$, $1\le j \le N_{in}$, $1\le \mu \le N$ and $k$ describing the convolutional kernel, $-(K-1)\le k \le K-1$, $K$ being the kernel size. Here, $U_{x, k \hat \mu}$ denotes the Wilson line of length $k$ along the lattice direction $\hat\mu$, i.e.
\begin{align} \label{eq:wilson_line_k}
    U_{x, k \hat \mu} = U_{x, \mu} U_{x+\hat\mu, \mu} \cdots U_{x+(k-1)\hat\mu, \mu}.
\end{align}
The trainable parameters are given by $\omega$. For compactness, one can also combine both layers into a single bilinear convolution layer (L-ConvBilin) as
\begin{align}
     W'^{l}_{x} = \sum_{k, \mu, m,n} \omega^{lmn}_{k,\mu} \,   W^m_{x} \, U_{x, k \hat \mu}  W^n_{x+ k \hat \mu}  U^\dagger_{x, k \hat \mu},
\end{align}
where the indices are analogous to the previous layers and $\omega$ is trainable. Before each layer, we extend the fields $W$ by including their Hermitian conjugates as additional channels, and we also add a single channel of unit elements, so that linear dependencies can be learned as well. 

To include the time dependence of the score function, we use a random Fourier feature embedding \cite{tancik2020fourier}, which is a common choice in diffusion model applications. Given the diffusion time $t$, one calculates 
\begin{align} \label{eq:fourier_embed}
    f^k(t)= \begin{cases}
        \sin(2\pi t w_h) & h<M/2 \\
        \cos(2\pi t w_{h-M/2}) & h \geq M/2
    \end{cases},
\end{align}
where $M$ is the embedding dimension and $w_h$ is a randomly initialized parameter vector of dimension $M/2$ that remains unchanged during training, $w_h \sim \mathcal{N}(0,r)$, with $r$ being some chosen scale (we set $r=3$). The resulting vector $f^h(t)$ is then passed through one fully connected dense layer giving $f^{l}_{\theta}(t)$. The time information is then added to the locally transforming variables between the bilinear-convolutional layers as
\begin{align}
     W'^{l}_{x} =  W^{l}_{x} + \mathds{1} \, f^l_{\theta}(t).
\end{align}
The number of channels of $ W$ in this step has to match the number of components of $f^l_\theta$. This time embedding retains the gauge equivariance of the L-CNN.

\begin{figure}
    \centering
    \includegraphics[width=0.7\linewidth]{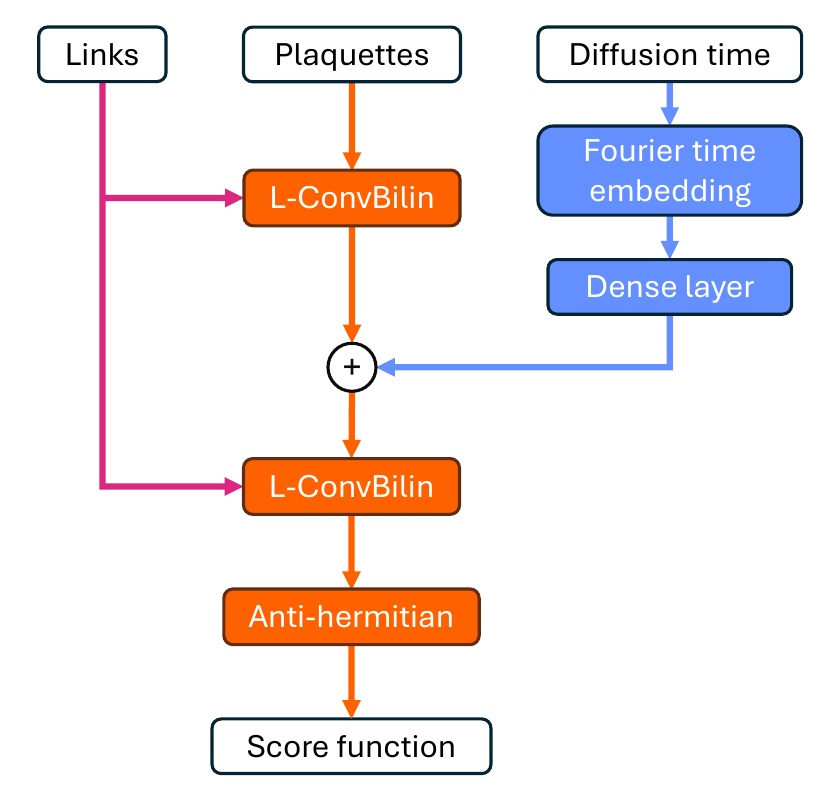}
    \caption{L-CNN network architecture.}
    \label{fig:network_base}
\end{figure}

\begin{figure}
    \centering
    \includegraphics[width=0.7\linewidth]{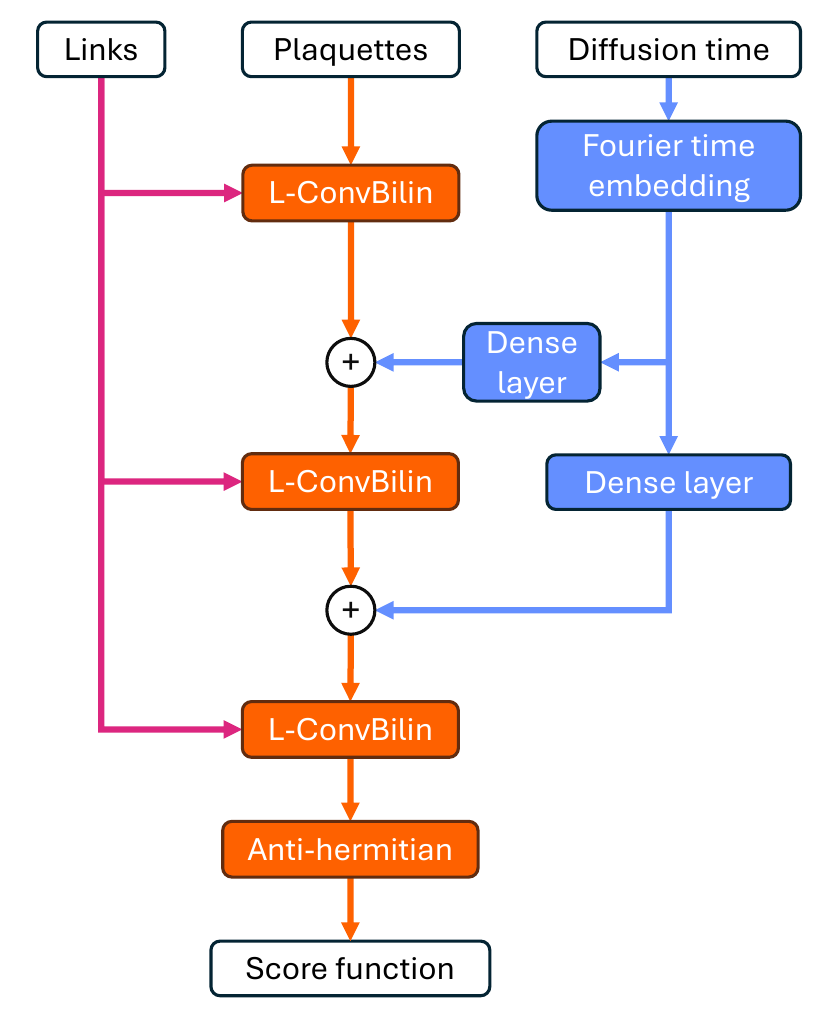}
    \caption{Larger L-CNN network architecture used in Fig.~\ref{fig:network_size}.}
    \label{fig:network_large}
\end{figure}

We have compared the performance of several parameter choices within the L-CNN architecture, see Section~\ref{sec:network_experiments}. The ideal architecture choice, which we use for all other numerical results, consists of two bilinear-convolution layers with a hidden channel size of $16$ and a convolution kernel size of $2$. Between the two layers, we add the time embedding, with an embedding dimension of $256$. At the end, we take $-i$ times the anti-Hermitian part of the output. In Fig.~\ref{fig:network_base} we visualize the chosen network architecture, and in Fig.~\ref{fig:network_large} we show the architecture of a larger network we test in Section~\ref{sec:network_experiments}.

\section{Metropolis-adjusted annealed Langevin algorithm} \label{sec:maala}
Besides solving the reverse SDE, we also follow a different approach for sampling field configurations, namely the Metropolis-adjusted annealed Langevin algorithm (MAALA) \cite{roberts1998optimal, Radivojevic2020}, applied to diffusion models in \cite{Zhu:2025pmw}. It connects the learned diffusion model with stochastic quantization (SQ)  \cite{Parisi:1980ys,Damgaard:1987rr}, in which the probability distribution is determined by an action $S$ as $p\propto \exp(-S)$. The score function is then $-D^a S$. In a diffusion model, we learn a time-dependent score function, which we now interpret as the negative gradient of some time-dependent action ${\cal S}_t$, see Eq.~(\ref{eq:St}). At fixed diffusion time, we can use this learned score to perform SQ sampling. 

We start with configurations drawn from the prior distribution at $t=T$. We then apply SQ updates to these configurations using the score function evaluated at $t=T-\Delta t$ for a given number of steps. The result is used as an initial condition for another SQ simulation at $t=T-2\Delta t$. This process is repeated until we reach $t=0$, where the score function is expected to correspond to minus the gradient of the action of our target theory. 

The SQ update is given by 
\begin{equation}
\label{eq:maala_gauge_sde}
 U^{(t)}_{(\tau+\epsilon)}=
\exp\!\left[ i \left( \epsilon \hat s_{\theta}(\mathbf{U}^{(t)}_{(\tau)}, t)
+\sqrt{2\epsilon} \, \hat \eta \right) \right]\,   U^{(t)}_{(\tau)},
\end{equation}
where $\tau$ corresponds to the Langevin time of SQ, $t$ is the diffusion time, and $\epsilon$ indicates here the time step in the SQ update. We suppressed the $x,\mu$ indices for clarity and remind the reader that the hat symbol denotes a contraction with the generators. 
At each time $t_i\in[T-\Delta t, T-2\Delta t, ... , \Delta t, 0]$, we solve Eq.~\eqref{eq:maala_gauge_sde} from $\tau=0$ to $\tau=\mathcal{T}$. 
The diffusion times $t_i$ do not need to be uniformly spaced. In practice, we use logarithmic time steps such that $\Delta t_i$ is much smaller for small $t$.
We note that we generate a single field configuration with a single chain of Langevin trajectories and solve many trajectories in parallel to obtain fully uncorrelated configurations. 

Under ideal conditions, in the limit of vanishing SQ discretization errors and assuming perfect training, samples generated using MAALA and those using the reverse SDE should asymptotically follow the same time-dependent probability distribution. The main advantage of MAALA lies in the ability to apply Metropolis adjustment to the Langevin trajectories to correct for any bias or error in the machine-learned score function. Given some field configuration $\mathbf{U}^{(t)}_{(\tau)}$, we obtain a proposal configuration $\tilde{\mathbf{U}}^{(t)}_{(\tau)}$ through Eq.~\eqref{eq:maala_gauge_sde}. We then choose $\mathbf{U}^{(t)}_{(\tau+\epsilon)}=\tilde{\mathbf{U}}^{(t)}_{(\tau)}$ with probability 
\begin{equation} \label{eq:p_metropolis}
    p_\text{accept} = \min \left\{ 1, \frac{p_0(\tilde{\mathbf{U}}^{(t)}_{(\tau)})}{p_0(\mathbf{U}^{(t)}_{(\tau)})} 
    \frac{q( \mathbf{U}^{(t)}_{(\tau)} | \tilde{\mathbf{U}}^{(t)}_{(\tau)}) }{q(\tilde{\mathbf{U}}^{(t)}_{(\tau)} | \mathbf{U}^{(t)}_{(\tau)}) } \right\},
\end{equation}
and $\mathbf{U}^{(t)}_{(\tau+\epsilon)}=\mathbf{U}^{(t)}_{(\tau)}$ otherwise. Here, $p_0$ is the target probability distribution, in our case determined by the Wilson action, and $q(\cdot |\cdot)$ is the transition probability of a single Langevin update step. The latter is defined by Eq.~\eqref{eq:maala_gauge_sde} to be 
\begin{equation}
    q(\tilde{\mathbf{U}}^{(t)}_{(\tau)} | \mathbf{U}^{(t)}_{(\tau)}) = q(\hat{\eta}^*),
\end{equation}
where $\hat{\eta}^*$ is such that Eq.~\eqref{eq:maala_gauge_sde} yields $\tilde{\mathbf{U}}^{(t)}_{(\tau)}$ given $\mathbf{U}^{(t)}_{(\tau)}$. Note that for compact groups there may be several $\hat{\eta}$'s allowing for the transition, but as we are performing updates with very small step sizes $\epsilon$, we only require the dominant contribution; all others are exponentially suppressed.
That dominant contribution is obtained by rewriting Eq.~\eqref{eq:maala_gauge_sde} to get 
\begin{align}
    \label{eq:metro_eta}
    \hat \eta^* &= \frac{1}{i \sqrt{2 \epsilon}} \ln(\tilde{{U}}^{(t)}_{(\tau)} {U}^{(t) \dagger}_{(\tau)}) - \sqrt{\frac{\epsilon}{2}} \hat s_{\theta}(\mathbf{U}^{(t)}_{(\tau)}).
\end{align}
As $\hat{\eta}$ is drawn from a normal distribution, 
\begin{equation}
    q(\hat{\eta}^*) = c \exp\left( -\sum_{x,\mu}\tr{\hat \eta^{*2}_{x,\mu} }\right),
\end{equation}
where $c$ is a normalization constant that drops out, we get
\begin{align}\label{eq:metro_acc}
    \frac{p_0(\tilde{\mathbf{U}})}{p_0(\mathbf{U})} \frac{q( \mathbf{U} | \tilde{\mathbf{U}}) }{q(\tilde{\mathbf{U}} | \mathbf{U}) } & = \exp\Big[ 
     - \left(S(\tilde{\mathbf{U}}) - S(\mathbf{U}) \right)   \nonumber \\
    & \quad + \sum_{x,\mu} \left( \tr{\hat \eta^{*2}}  - \tr{\hat \eta'^{*2}} \right)
    \Big],
\end{align}
with $\hat \eta'^{*}$ defined analogously to Eq.~\eqref{eq:metro_eta} but with $\mathbf{U}$ and $\tilde{\mathbf{U}}$ exchanged. In Appendix \ref{app:metropolis}, we present a slightly different sampling method which requires no approximations. In practice, we have used the matrix logarithm (with the principal branch) for our numerical simulations, which gives identical results for small time steps $\epsilon$.

We apply Metropolis adjustment only to the last $5\%$ of Langevin trajectories, i.e., at small diffusion times $t$, when the time-dependent action and the learned score are expected to be close to the target action and minus its gradient.  

The second advantage of MAALA is the ability to use \emph{score rescaling}, that is, sampling at inverse couplings different from the coupling of the training data. If a score function is trained at some $\beta_0$, we assume
\begin{equation}
s_{\theta}^a(U, t\to 0) \sim -D^aS(U; \beta_0),
\end{equation}
with $S(U; \beta_0)$ the Wilson action at gauge coupling $\beta_0$. Since the Wilson action is proportional to the coupling, rescaling the score function by a factor of $\beta/\beta_0$ yields 
\begin{equation}
\frac{\beta}{\beta_0} s_{\theta}^a(U, t\to 0) \sim -D^aS(U; \beta). 
\end{equation}
Since the score function describes a continuous map for $t\in [0,T]$ between the target action at $\beta_0$ and the Haar measure, we can scale the learned score by $\beta/\beta_0$ at all $t$ to get a continuous map between the target action at $\beta$ and the Haar measure  (since the Haar measure corresponds to $\beta=0$, rescaling has no effect here).
Through this score rescaling, we can, in principle, sample at arbitrary couplings with MAALA, even when we have only a training ensemble at a single coupling. We note that the evolution of the probability distribution obtained through score rescaling from $\beta_0$ to $\beta$ will, in general, be different from the one given by the forward diffusion process carried out directly at the $\beta$ value. The rescaled score does not correspond to a known forward process; therefore, the reverse SDE is also unavailable. This is illustrated in Fig.~\ref{fig:maala_reverse} below, where a clear (and expected) mismatch between the forward and the backward MAALA processes is visible for $\beta\neq \beta_0$.

\section{Benchmarking, optimization and numerical results}  \label{sec:numerical_results}

For our numerical simulations, we focus on two-dimensional U(2) gauge theory.
The reason is that the topological susceptibility, a notoriously difficult observable to obtain, is trivially zero in two-dimensional SU(2), but not in U(2). Results for SU(2) can be found in our previous work \cite{Aarts:2026zzr}.
We train the L-CNN network using $100\,000$ U(2) configurations generated with Hybrid Monte Carlo (HMC) at $\beta_0=2$ on a $16^2$ lattice. 
Exact analytical results on a finite volume are available and presented in Appendix~\ref{app:analytical_solution}. 
In most figures the lattice size for generation is $16^2$ as well; if not, this is clearly indicated.

Specific observables we look at are traced square Wilson loops $\langle W_{n \times n} \rangle$ and the topological susceptibility $\chi_\mathrm{top}$. 
The traced Wilson loops are defined by
\begin{align}
    W_{x, n \times n} =
    \frac{1}{N_c} \tr{ U_{x,n \hat \mu } U_{x + n \hat \mu, n \hat \nu } U^\dagger_{x + n \hat \nu, n \hat \mu } U^\dagger_{x,n \hat \nu }},
\end{align}
where we used the Wilson line of length $n$ as defined in Eq.~\eqref{eq:wilson_line_k}. We average over all $n \times n$ loops. The topological susceptibility is defined as
\begin{align}
    \chi_\mathrm{top} = \frac{\langle Q^2 \rangle}{L^2}, \qquad Q = \frac{1}{2\pi} \sum_x \arg (\det P_{x,\mu \nu}) \in \mathbb Z,
\end{align}
with $Q$ denoting the topological charge. 

\subsection{Reverse SDE sampling}

\begin{figure}
    \centering
    \includegraphics[width=0.95\linewidth]{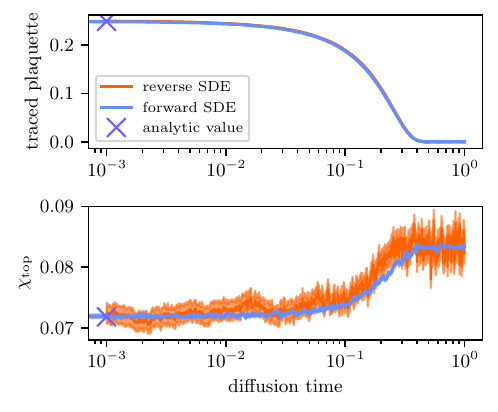}
    \caption{Comparison of the expectation value of the traced plaquette $\langle\tr{U_P}\rangle$ and the topological charge susceptibility $\chi_{\rm top}$ along the forward diffusion process and the learned reverse diffusion process for U(2). Training and generation are both carried at $\beta_0=2$ on a $16^2$ lattice. Crosses denote the exact (analytical) values.
     }
    \label{fig:reverse_sde_process}
\end{figure}

We solve the reverse SDE~\eqref{eq:reverse_gauge} for initial configurations drawn from the Haar measure, using the trained network as a score function to generate $5000$ configurations.
In Fig.~\ref{fig:reverse_sde_process}, we compare the evolution of the expectation value of the traced plaquette and the topological charge susceptibility along the reverse SDE flow with the one obtained by applying the forward SDE to $50\,000$ HMC samples. 
We used more HMC samples, to ensure statistical fluctuations of the benchmark ensemble are under control. 
In the generative model, no Metropolis correction is applied. The topological susceptibility is considerably noisier than the traced plaquette, as expected. 
We observe that the forward and backward paths agree very well, connecting 
the values in the target ensemble with the values evaluated on the Haar measure ($\beta=0$), where traced Wilson loops vanish and $\chi_{\rm top}=1/12$.

\begin{figure}
    \centering
    \includegraphics[width=0.95\linewidth]{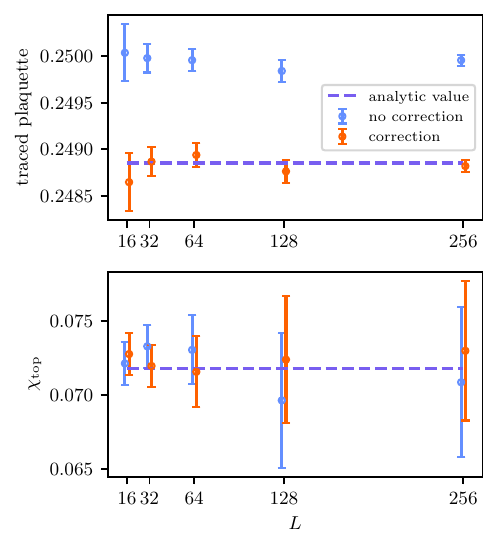}
    \caption{The same observables as in the preceding figure, as a function of lattice volume at $\beta_0=2$. The ensembles are generated using the reverse SDE with the network trained at $\beta_0=2, L=16$, but employed on larger lattices (extrapolation).
    The remaining bias in the plaquette is corrected by applying a few standard SQ steps with Metropolis adjustment. 
    For $\chi_{\rm top}$ the statistical uncertainty is larger and the (un)corrected results are consistent within the error with the analytical prediction.
    Note that the data points have been slightly shifted along the horizontal axis for clarity.
    }
    \label{fig:reverse_sde_size_extrapolation}
\end{figure}

The crosses in Fig.~\ref{fig:reverse_sde_process} indicate the exact analytical results on a finite volume. For the plaquette, a small bias is present, which is not visible in  Fig.~\ref{fig:reverse_sde_process}, but can be seen in Fig.~\ref{fig:reverse_sde_size_extrapolation}, where we show the results of extrapolating to larger lattices, for the same observables. At the largest lattices, $L=128$ and $L=256$, we generated only $500$ configurations.
We emphasise that the ensembles are generated using the network trained at $L=16$, but employed on larger lattices. The bias can be corrected by applying a small number of standard SQ steps with Metropolis adjustment to the reverse SDE results. 
For $\chi_{\rm top}$ the statistical uncertainty is larger and the (un)corrected results are consistent within the error with the analytical prediction.
Importantly, the extrapolation to larger lattices does not significantly increase errors.

\subsection{MAALA sampling}

\begin{figure}
    \centering
    \includegraphics[width=0.95\linewidth]{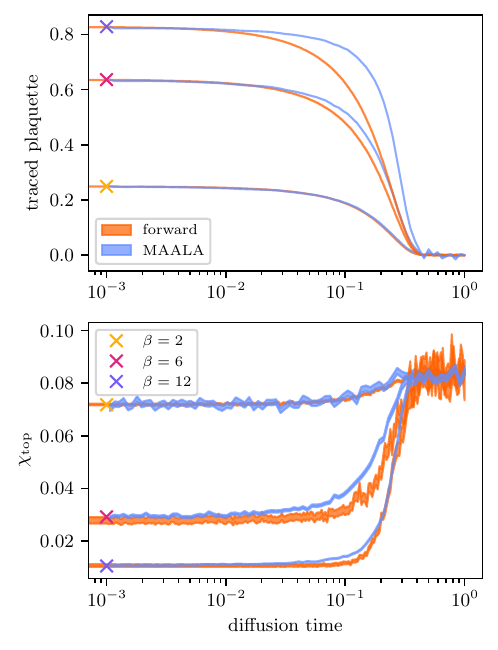}
    \caption{Comparison of the evolution of $\langle\tr{U_P}\rangle$ and $\chi_{\rm top}$ obtained using the forward diffusion process applied to HMC samples at three $\beta$ values and the generative MAALA algorithm using score rescaling, starting from a neural network trained at $\beta_0=2$, $L=16$. 
    Crosses denote the analytical values at the respective $\beta$ values.  
    When $\beta\neq \beta_0$, the forward and backward paths as described here are different, but the start and endpoints are  determined by the target $\beta$ values and the Haar measure, as expected.  
    }
    \label{fig:maala_reverse}
\end{figure}

Having shown that sampling using the reverse SDE works well for all considered lattice sizes, 
we now continue with generating configurations using the Metropolis-adjusted annealed Langevin algorithm (MAALA). 
In Fig.~\ref{fig:maala_reverse} we compare the generative process of MAALA with the forward diffusion process applied to HMC samples. As in the case of the reverse SDE, the MAALA generative process agrees with the forward diffusion process on which the neural network was trained ($\beta_0=2$). When we apply score rescaling to generate at other $\beta$ values, we find that the generative process and the forward diffusion process applied to HMC samples at the corresponding $\beta$ value follow different paths through the space of distributions. This is expected, as score rescaling is only expected to be correct at $t\to 0$, while at larger diffusion times $t > 0$, the forward process defines a complicated time-dependent score function at a given $\beta$ that cannot be described by mere rescaling. We give further evidence of this in Section~\ref{sec:time-dep-action}.

\begin{figure}
    \centering
    \includegraphics[width=0.95\linewidth]{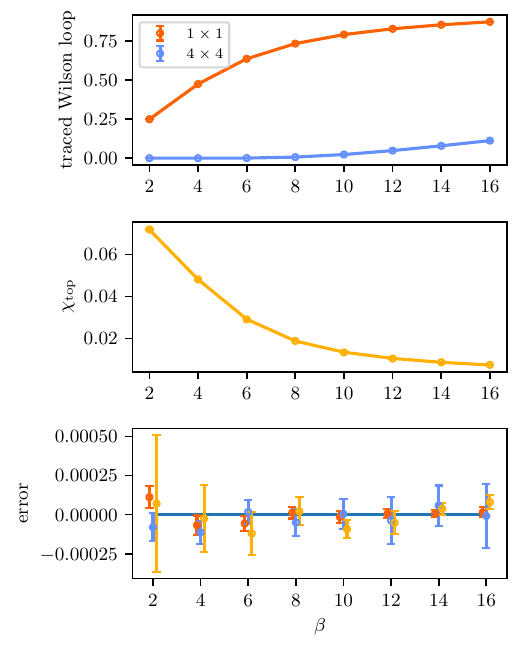}
    \caption{Traced plaquette and $4\times 4$ Wilson loop (above) and $\chi_{\rm top}$ (middle) versus $\beta$, computed using ensembles generated with MAALA (circles) and the analytic values (lines). The network was trained at $\beta_0=2$; larger values of $\beta$ are obtained via score rescaling. The bottom pane contains the difference between the generated and exact values; data points have been slightly shifted horizontally for clarity.}
    \label{fig:beta_extrapolation}
\end{figure}

For our high-precision extrapolation experiments, we train ten networks with identical hyperparameters, but different randomly initialized training weights. We then take the average over the ten results, obtained from $10\,000$ configurations each, using the DerSimonian-Laird analysis \cite{DerSimonian:1986aaa, DerSimonian:2015aaa, Cochran:1954aaa} (see also Appendix \ref{app:errors}).
In Fig.~\ref{fig:beta_extrapolation} we show the traced plaquette and $4\times 4$ Wilson loop and the topological susceptibility as a function of $\beta$, obtained via score rescaling; a network is trained at $\beta_0=2$ and all other ensembles are generated by multiplying the trained score with $\beta/\beta_0$. Lines indicate the analytical values. The bottom pane contains the difference between the generated and exact values.
Excellent agreement up to the highest considered $\beta=16$ is observed. 

\begin{figure}
    \centering
    \includegraphics[width=0.95\linewidth]{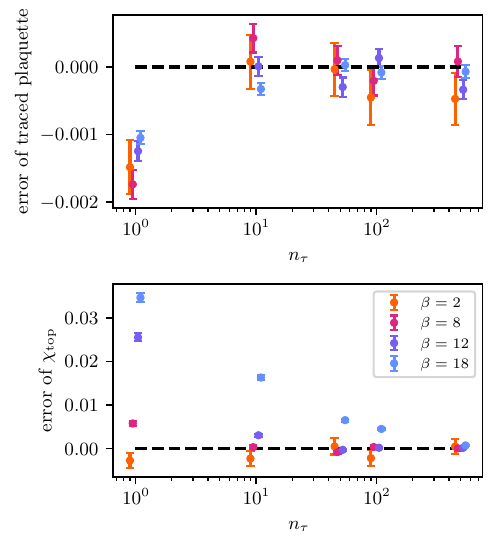}
    \caption{Comparison of the error for $\langle\tr{U_P}\rangle$ and $\chi_{\rm top}$ at four $\beta$ values obtained by running MAALA with different Langevin trajectory lengths ${\cal T}=\epsilon n_\tau$, with fixed $\epsilon=0.005$. For $\chi_{\rm top}$, longer Langevin trajectories are required at larger $\beta$ values to obtain accurate results. }
    \label{fig:langevin_length_over_beta}
\vspace*{0.4cm}
    \centering
    \includegraphics[width=0.95\linewidth]{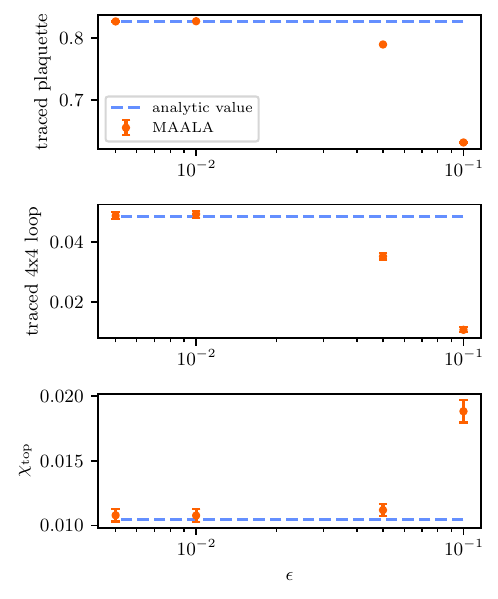}
    \caption{
    As in the preceding figure, varying now the step size $\epsilon$ at fixed ${\cal T}$ for $\beta=12$. 
    Discretization artifacts are negligible for $\epsilon \lesssim 0.01$.
    }
    \label{fig:vary_langevin_stepsize}
\end{figure}

Next we investigate the effect of changing the parameters of the Langevin trajectories within the MAALA sampling scheme. 
First we change the length of the Langevin trajectory $\mathcal T$ at fixed step size $\epsilon=0.005$ by increasing the number of steps $n_\tau$, where $\mathcal T = n_\tau \epsilon$. 
The length of each trajectory is held constant along diffusion time $t$.
In Fig.~\ref{fig:langevin_length_over_beta}, the dependence of the error of the observables on $n_\tau$ is shown. Especially for $\chi_{\rm top}$, longer Langevin trajectories are required at larger $\beta$ values to obtain accurate results.
Second, we vary the Langevin step size $\epsilon$ at fixed Langevin trajectory length $\mathcal T$, to investigate the step size required to solve the Langevin SDE in MAALA and avoid discretization errors. The results are shown in Fig.~\ref{fig:vary_langevin_stepsize}. We observe that discretization artifacts are negligible for $\epsilon \lesssim 0.01$. 
For all other results in this paper, we used a step size of $\epsilon=0.005$.

\begin{figure}
    \centering
    \includegraphics[width=0.95\linewidth]{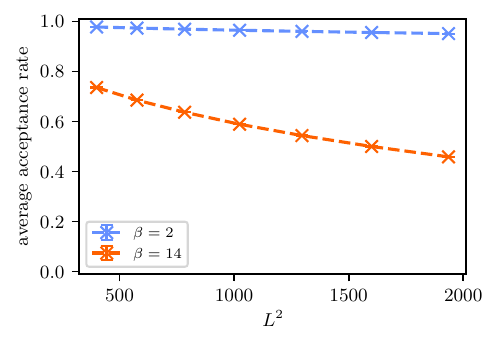}
    \caption{Dependence of the Metropolis acceptance rate in the MAALA sampling process when extrapolating to larger volumes $L^2$. The model was trained on a  lattice with $\beta_0=2$ and $L = 16$; we show results when this model is applied to lattices of size up to $L = 44$, for $\beta=2$ and $14$. }
    \label{fig:volume_scaling}
\end{figure}

We further investigate the scaling behavior of our sampling method as we extrapolate to larger lattices. To use a single metric for sampling quality, we use the average acceptance rate of the Metropolis adjustment applied during the last $5\%$ of the MAALA update steps. Note, however, that this metric cannot be directly related to quantities like the effective sample size. In Fig.~\ref{fig:volume_scaling}, we show how the acceptance rate changes over the lattice volume for two different inverse couplings. Similar to sampling using the reverse SDE in Fig.~\ref{fig:reverse_sde_size_extrapolation}, the agreement between the measured observables and the analytic values does not significantly worsen during the extrapolation to larger lattices.

\subsection{Network architectures} \label{sec:network_experiments}

We turn to the performance of different L-CNN architectures. To simplify the comparison, we again use the average Metropolis acceptance rate in the generative process. We train all networks at $\beta_0=2$ and investigate the acceptance rates when generating at $\beta=18$, where the generative task is especially hard and the most noticeable performance differences are expected. For each architecture, we train five networks with different random initializations and present the five resulting acceptance rates. We use the same number of gradient descent steps during training for all architectures.
The properties of the four different networks we compare are shown in Table~\ref{tab:network_sizes}.
The base model has, as described in Section~\ref{sec:lcnn}, two L-ConvBilin layers with a hidden channel size of $16$ and a kernel size of $2$ for both layers. 
For the larger model, we add a third L-ConvBilin layer, also with hidden channel size of $16$ and kernel size $2$. For the smaller model, we have two layers, but with a hidden channel size of $4$, and the kernel of the second layer has size $1$. Finally, for the tiny model we use only a single L-ConvBilin layer, with the time embedding being applied after the L-ConvBilin. 

\begin{table}
    \centering
    \begin{ruledtabular}
    \begin{tabular}{l|rrrr}
         & Base & Large & Small & Single layer \\ \hline
        Number of layers & 2 & 3 & 2 & 1 \\ 
        Hidden channels & 16 & 16,\,16 & 4 & - \\ 
        Kernel sizes & 2,\,2 & 2,\,2,\,2 & 2,\,1 & 2 \\ 
        Trainable parameters & 15\,266 & 104\,386 & 1\,322 & 580 \\
    \end{tabular}
    \end{ruledtabular}
    \caption{Details of four L-CNN architectures. The number of output channels of the time embedding layer is adapted to each architecture.}
    \label{tab:network_sizes}
\end{table}

\begin{figure}
    \centering
    \includegraphics[width=0.95\linewidth]{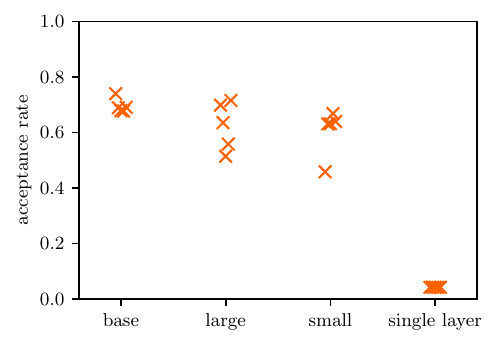}
    \caption{Comparison of the acceptance rate when generating at $\beta=18$, $L=16$ for four different L-CNN architectures (see Table~\ref{tab:network_sizes}). For each architecture, five models are trained independently.
        }
    \label{fig:network_size}
\end{figure}

The results are shown in Fig.~\ref{fig:network_size}. We find that while the single-layer model fails completely, the three other architectures achieve similar results. The base model slightly outperformed the other architectures; we have chosen this model for all other computational experiments. 

\begin{figure}
    \centering
    \includegraphics[width=0.95\linewidth]{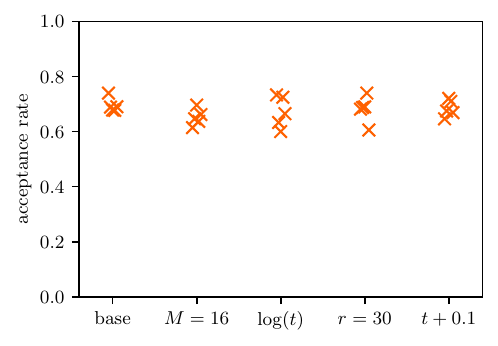}
    \caption{As in the preceding figure, using different time embedding hyperparameters within the L-CNN.}
    \label{fig:time_embedding}
\end{figure}

Next, we compare different hyperparameter choices for the time embedding in Fig.~\ref{fig:time_embedding}. For the embedding dimension $M$, we compare our default choice of $256$ to a smaller choice of $16$. Further, we change the embedding scale $r$ from $3$ to $30$, and shift the time $t$ by an increment of $0.1$ such that we do not reach zero in Eq.~\eqref{eq:fourier_embed}. Finally, we re-parametrise the time embedding (denoted as ``$\log(t)$'' in Fig.~\ref{fig:time_embedding}): instead of $t$ we use $t' = (\log(t)-10^{-3})/\log(10^{-3})$, giving more weight to the early times of the diffusion process, where one might expect more significant changes in the score function. Our results show that none of these changes amount to a significant difference in performance, suggesting that the time dependence of the score function is simple enough to learn. Thus, a hyperparameter optimization of the time embedding is unnecessary.

\subsection{Comparison to stochastic quantization}

We compare our diffusion sampler to conventional Langevin methods, the simplest being standard stochastic quantization, where we use the Wilson action at the target $\beta$ value in the drift term of the Langevin equation. Further, we test an annealed Langevin method: we use the Wilson action in the Langevin equation but linearly increase $\beta$ from zero to the target value along the Langevin trajectory. In both cases, we also apply Metropolis adjustment towards the end of the trajectory, analogous to our MAALA implementation. For all three methods, we use the same number of Langevin update steps $n_\tau$ with the same step size $\epsilon=0.005$.

\begin{figure}
    \centering
    \includegraphics[width=0.95\linewidth]{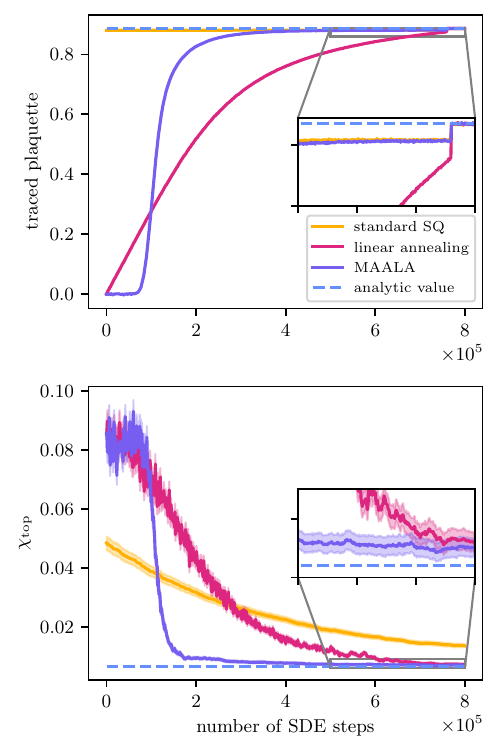}
    \caption{Comparison of the generative process at $\beta=18$ and $L=16$ using different Langevin-based sampling methods. Standard SQ uses the Wilson action at $\beta=18$, and linear annealed SQ also uses the Wilson action, but with a linearly increasing $\beta$, reaching the target value after $8\times 10^5$ steps. MAALA uses the learned score function, trained at $\beta_0=2$. The insets show a blow up of the final part of the evolution, including the switching on of the Metropolis adjustment. 
    }
    \label{fig:langevin_comparison_short}
\end{figure}

\begin{figure}
    \centering
    \includegraphics[width=0.95\linewidth]{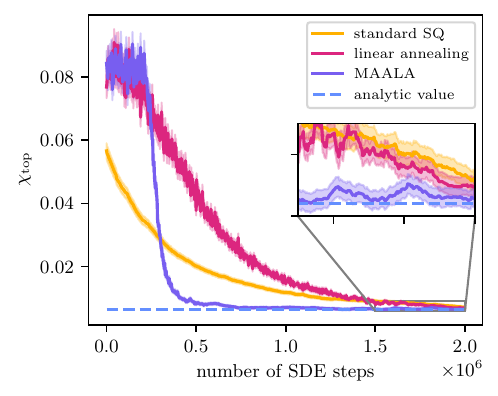}
    \caption{As in the preceding figure, with a larger number of SDE steps. }
    \label{fig:langevin_comparison_long}
\end{figure}

In Fig.~\ref{fig:langevin_comparison_short}, we plot the comparison for a target coupling of $\beta=18$. All methods are able to correctly reproduce the average plaquette due to the Metropolis adjustment, which is responsible for the jump towards the end of the generative process (top pane of Fig.~\ref{fig:langevin_comparison_short}). 
In SQ, in which observables depend linearly on the step size, the plaquette thermalizes very quickly, but the expectation value is slightly too low, which is cured by the Metropolis adjustment. In annealed Langevin, $\beta$ increases linearly and reaches the target value after $8\times 10^5$ steps. This explains the slow but steady rise to the analytic value. 
Methods that use annealing perform significantly better for the topological susceptibility, with MAALA stabilizing quickly (bottom pane of Fig.~\ref{fig:langevin_comparison_short}). However, none of the methods can reach the analytic value with the chosen number of steps. When increasing the total number of steps, the performance difference decreases and the gap between measured and analytical results closes for all methods, see Fig.~\ref{fig:langevin_comparison_long}. This is to be expected, as all sampling methods in this comparison are asymptotically exact.

\subsection{Structure of the time-dependent action}
\label{sec:time-dep-action}

In the previous subsection, it was shown that the MAALA sampling process may resemble an annealed Langevin process with a time-dependent coupling. Given the simplicity of two-dimensional lattice gauge theory, it is reasonable to check whether the forward process simply induces such a time-dependent coupling $\widetilde\beta(t)$ without changing the structure of the Wilson action, i.e., whether the integral over the time-dependent (but unknown) probability distribution in Eq.~(\ref{eq:St}) can be replaced by the partition function
\begin{align}
    \widetilde{\mathcal Z}(t) = \int \mathcal D \mathbf U\, \exp\left(-\frac{\widetilde\beta(t)}{N_c} \sum_{P} \mathrm{Re}\, \mathrm{Tr}\, [\mathds{1} - U_P] \right),
\end{align}
where the sum runs over all positively oriented plaquettes $P$. In two dimensions, in the infinite-volume limit (or with open boundary conditions), it is well known \cite{Bonati:2019ylr} that the functional integral can also be carried out over plaquettes directly and therefore factorizes, such that
\begin{align}
    \widetilde{\mathcal Z}(t) = \prod_P \int \mathcal D U_P\, \exp\left(-\frac{\widetilde{\beta}(t)}{N_c} \mathrm{Re}\, \mathrm{Tr}\, [ \mathds{1}- U_P] \right). \label{eq:factorized_Z}
\end{align}
This simplification is only possible if the action contains only plaquette terms and no larger Wilson loops. 

The special structure of Eq.~\eqref{eq:factorized_Z} leads to factorized expectation values of traced plaquettes $U_P \neq U_P'$, 
\begin{align} \label{eq:f}
    \langle \tr{U_P} \tr{U_{P'}}\rangle_{\widetilde\beta(t)} = \left(\langle \tr{U_P} \rangle_{\widetilde\beta(t)} \right)^2,
\end{align}
for all time. On the other hand, in Section \ref{sec: analytical_diffusion} we have shown analytically that the forward diffusion process entangles the links and such a factorization is not possible. Here we test this aspect numerically for the forward and backward processes.

\begin{figure}
    \centering
    \includegraphics[width=0.95\linewidth]{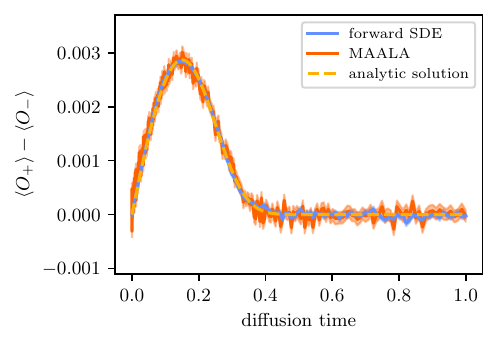}
    \caption{Difference between $\langle O_{+} \rangle$ and $\langle O_{-} \rangle$ on a lattice with $\beta=2, L=16$, during the forward diffusion process, both analytically and numerically, and during the generative MAALA process. A non-zero difference indicates that the underlying action is not merely the Wilson action with some time-dependent coupling.}
    \label{fig:connected_plaq_difference}
\end{figure}

To do so, we compute the observables $O_\pm$ (see Eq.~\eqref{eq:plaq_correlators}) along the forward and MAALA backward processes.
These observables, repeated here for clarity, are 
\begin{align}
    \begin{split}
    O_+ &= \frac{1}{N_c^2} \Re\, \tr{U_P} \tr{U_{P'}}, \\
    O_- &= \frac{1}{N_c^2} \Re\, \tr{U_P} \tr{{U_{P'}}^\dagger},
    \end{split} 
\end{align}
where $U_P$ and $U_P'$ share one link. If the evolution is described by the Wilson action with a time-dependent gauge coupling, $\langle O_+\rangle=\langle O_-\rangle$ at all times, see Eq.~(\ref{eq:f}). 
Analytical expressions for the evolution under the forward diffusion process are derived in Appendix \ref{app:eff_action} and yield $\langle O_+\rangle \neq\langle O_-\rangle$ when $t>0$. The equality holds at $t=0$ and again at the end of the diffusion process, i.e., for the Haar measure. 

In Fig.~\ref{fig:connected_plaq_difference} we show the time dependence of $\langle O_+\rangle - \langle O_-\rangle$ at $\beta=2$, 
via the analytical solution of the forward diffusion process, 
the numerical evaluation of the diffusion process starting from an HMC ensemble on an $L=16$ lattice, 
and the backward evolution obtained with the generative MAALA process. 
We find excellent agreement between all three results. 
We note that the difference increases linearly in $t$ and hence the time-dependent action deviates from a Wilson action immediately. At large $t$, the difference vanishes exponentially, as the distribution  approaches the Haar measure. 
Since at intermediate times indeed $\langle O_+\rangle \neq\langle O_-\rangle$, 
the diffusion process induces a more complicated lattice action than the Wilson action with a time-dependent coupling. 
Similarly, the L-CNN encoding the negative gradient of this effective action cannot be represented by plaquette terms only.

\subsection{(Non-)conservation of Creutz ratios}

\begin{figure}
    \centering
    \includegraphics[width=0.95\linewidth]{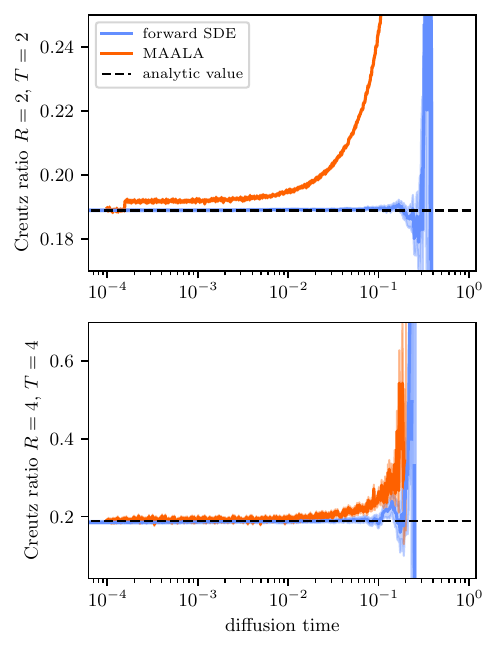}
    \caption{Creutz ratios $\chi(2,2)$ (top) and $\chi(4,4)$ (bottom) in the fundamental representation for U(2) at $\beta=12$ on a lattice with $L=16$ along the forward diffusion process applied to HMC configurations (forward SDE) and along the backward generative MAALA process, using the network trained at $\beta_0=2$ with score rescaling. The dashed line is the exact value, which is conserved analytically under forward diffusion. 
    } 
    \label{fig:creutz_ratio}
\end{figure}

In Section \ref{sec: analytical_diffusion} we showed analytically that Creutz ratios are exactly conserved under the forward diffusion process, see Eq.~(\ref{eq:Creutz}). This leads to a puzzle how they can be reconstructed under the generative process, starting from a trivial prior. 
In Fig.~\ref{fig:creutz_ratio} we show the evolution of Creutz ratios in the fundamental representation for $R=T=2$ and $R=T=4$. Analytically, these are exactly conserved. 
However,  as the conservation relies on the cancellation of noisy terms in linear combinations of $\log \bra W_{R,T}\ket_\tau$, in the numerical implementation of the forward process Creutz ratios suffer from incomplete cancellations, leading to large fluctuations and deviations from the expected value when $t\gtrsim0.1$. 

Concerning the backward process, Wilson loops initially vanish, due to the choice of prior. Hence during the backward evolution, the Creutz ratios are not conserved but need to be rebuilt during the generative process. This is also demonstrated in Fig.~\ref{fig:creutz_ratio}. We observe that the switching on of the Metropolis adjustment is visible at the end of the backward process and that it is required to reach the correct value for the target theory. 
We note here that the presence of noisy cancellations and the rebuilding of, in principle, conserved quantities is analogous to what happens to higher-order cumulants in the variance-expanding scheme on flat manifolds \cite{Aarts:2024rsl}. 

\section{Conclusions and outlook}

In this companion paper to \cite{Aarts:2026zzr}, we have presented a thorough extension of score-based generative diffusion models to non-abelian lattice gauge theory. 
We have separated our analysis for non-abelian groups in the forward diffusion process, which does not rely on machine learning, and the backward generative process, for which we demonstrated explicitly that learning is required.
For the forward evolution, we provided a complete analytical analysis of the time dependence of rectangular Wilson loops and similar observables, which can be used to benchmark the numerical implementation. For the backward process, we demonstrated that the loss function is gauge invariant provided the score is learned by a gauge-equivariant network, such as the lattice convolutional neural network (L-CNN). The dependence on the choice of L-CNN architectures is studied. 
Several theoretical subtleties specific to the non-abelian nature are discussed and we have investigated the validity of some necessary approximations. 

Turning to the numerical benchmarks, we have compared a number of generative approaches: the standard diffusion model approach, using reverse stochastic dynamics, with and without a small corrective step, the Metropolis-adjusted annealed Langevin algorithm (MAALA), which includes additional Langevin-type dynamics and is asymptotically exact, and standard stochastic quantization and annealed evolution which do not require machine learning. 
A corrective Metropolis adjustment was shown to be required in all cases, but only needs to be applied towards the end of the backward process.
A powerful result is that a model can be trained at one value of the gauge coupling and on one lattice volume, but the generation of new ensembles can then be generalized to different gauge couplings and larger volumes, without an increase in bias. 
Specifically, we have trained models for U(2) gauge theory in two dimensions on a single HMC generated ensemble at $\beta=2$ and $L=16$, and generated new ensembles with couplings and volumes up to $\beta=18$ and $L=256$.

By construction, the method generates statistically independent configurations and therefore exhibits no topological freezing in the conventional sense.
For all observables we considered, results consistent with the exact values were obtained at all lattice sizes and couplings we analyzed. However, for the largest gauge couplings considered ($\beta \sim 18$), we found that significantly longer Langevin trajectories are required to obtain accurate results.
This effect may be analogous to critical slowing down. On the other hand, when extrapolating to larger lattices, we find only a moderate decrease of the Metropolis acceptance rate with volume. Our comparison of MAALA to standard methods demonstrates the advantage of using the learned time-dependent score function. Additionally, we show both analytically and numerically that the score function is distinct from the gradient of the Wilson action multiplied by a time-dependent factor, establishing unambiguously that the learned time-dependent effective action is more complicated than a simple modification of the Wilson action.

Conceptually, our method directly extends to four-dimensional SU(3) gauge theory, as we have not made use of any specific assumptions regarding the dimensionality of the lattice or the gauge group, which can be SU($N_c$) or U($N_c$). The next step is therefore to investigate if practical application to the SU(3) theory in four dimensions is feasible, and if performance gains compared to standard methods can be achieved. Recently, several training and sampling strategies have been introduced, partially for non-abelian lattices \cite{Komijani:2026lan} and partially for other lattice theories \cite{Tomiya:2026tbc,Sexty:2026aym}. It will be interesting to see how they compare on larger non-abelian lattices. 
Another natural next step would be to include fermions, which has already been done for flow-based methods \cite{Abbott:2022zhs, Abbott:2022hkm} and for the Schwinger model also with diffusion models \cite{Vega:2026lrm}.
It would also be interesting to train diffusion models on ensembles stemming from more complicated lattice actions, e.g.,~fixed-point actions \cite{Holland:2024muu, Holland:2025fsa}, which are computationally costly to use in traditional Monte Carlo simulations.

\begin{acknowledgments} 
We thank Biagio Lucini and Kai Zhou for insightful discussions.
We thank the DEEP-IN working group at RIKEN iTHEMS for support in the preparation of this paper.
GA was supported by STFC grant ST/X000648/1 and by a Royal Society Leverhulme Trust Senior Research Fellowship, and is currently supported by the W\"ubben Stiftung Wissenschaft. 
DEH is supported by the UKRI AIMLAC CDT EP/S023992/1. 
AI, DM and TR acknowledge funding from the Austrian Science Fund (FWF) 10.55776/PAT3667424.
LW is supported by JSPS KAKENHI Grant No.~25H01560, and JST-BOOST Grant No.~JPMJBY24H9. LW and QZ are also supported by the RIKEN-TRIP initiative (RIKEN-Quantum). 
WW and QZ are supported in part by the Natural Science Foundation of China under Grants No.~12125503 and No.~12305103. 
The computational results have been achieved using the Austrian Scientific Computing (ASC) infrastructure.
\end{acknowledgments}

\section*{Data availability}

\paragraph*{Research Data and Code Access}
The code used for this work is available from Ref.~\cite{ranner_2026_24131871}. Research data is available upon request.

\paragraph*{Open Access Statement}
For the purpose of open access, the authors have applied a Creative Commons Attribution (CC BY) license to any Author Accepted Manuscript version arising.

\section*{LLM disclosure}
Anthropic Claude Opus 5.5 and Sonnet 5 assisted with small parts of code development and analytical derivations, and with manuscript polishing. 
Figure \ref{fig:OpOm} was made by Claude Opus 5.5. All other figures have been proudly produced by the authors. 
The authors directed the use of AI, fully understood and verified all code and derivations, and take full responsibility for the content of this work.

\appendix

\section{Properties of group derivatives} \label{app:group_derivative}

We consider a differentiable scalar function $f$ on the group $\GG$ (either SU$(N_c)$ or U$(N_c)$). For simplicity we restrict the discussion to functions $f(U)$ of a single argument $U$. The generalization to a lattice of links $U_{x,\mu}$ is straightforward.

The group derivative $D^a f$ is defined tangential to the group manifold via
\begin{align} \label{eq:group_der_def1}
    D^a f(U) = \frac{\dd }{\dd \epsilon} f(e^{i \epsilon T^a} U) \rvert_{\epsilon = 0},
\end{align}
or equivalently
\begin{align} \label{eq:group_der_def2}
    D^a f(U) = \lim_{\epsilon \rightarrow 0} \frac{1}{\epsilon} [f(e^{i \epsilon T^a} U) - f(U)].
\end{align}
The derivative is linear
\begin{align}
    D^a[\alpha g(U) + \beta h(U)] = \alpha D^a g(U) + \beta D^a h(U),
\end{align}
for $\alpha, \beta \in \mathbb C$ and two differentiable functions $g$ and $h$. Group derivatives also follow the product rule
\begin{align}
    D^a [g(U) h(U)] = \left[ D^a g(U) \right] \, h(U) + g(U) \, D^a h(U),
\end{align}
and satisfy the usual rule of integration by parts
\begin{align} \label{eq:group_der_int_parts}
    \int \DD U \, g(U) D^a h(U) = - \int \DD U \, \left[ D^a g(U) \right] \, h(U). 
\end{align}
This follows immediately from the left-invariance of the Haar measure,
\begin{align}
    \int \DD U f(U) = \int \DD U \, f(V U), \qquad \forall \, V \in \GG.
\end{align}
Setting $V = e^{i \epsilon T^a}$ we can rearrange the above equation to obtain
\begin{align}
    \int \DD U \, [f(e^{i \epsilon T^a} U) - f(U)] = 0,
\end{align}
which, upon taking the limit $\epsilon \rightarrow 0 $, yields
\begin{align}
    \int \DD U \, D^a f(U) = 0.
\end{align}
With $f(U) = g(U) \, h(U)$ one obtains Eq.~\eqref{eq:group_der_int_parts}.

From Eq.~\eqref{eq:group_der_def2}, we obtain a useful relation
\begin{align} \label{eq:group_der_taylor1}
    f(e^{i \varepsilon T^a} U) = f(U) + \epsilon D^a f(U) + \mathcal{O}(\epsilon^2).
\end{align}
One can compare this to a direct Taylor expansion in terms of complex matrix elements $U_{ij}$ (discarding terms of order $\epsilon^2$)
\begin{align}
    f(e^{i \varepsilon T^a} U) &= f((1+i\epsilon T^a) U) + \mathcal O(\epsilon^2) \nonumber \\
    &\approx f(U) + \delta U_{ij} \frac{\partial f}{\partial U_{ij}} + \delta \bar U_{ij} \frac{\partial f}{\partial \bar U_{ij}},
\end{align}
where $\delta U = i \epsilon T^a U$ is a small perturbation. With $\delta \bar U_{ij} = - i \epsilon \bar T^a_{ik} \bar U_{kj} = - i \epsilon  (U^\dagger)_{jk} T^a_{ki} $ we get 
\begin{align} \label{eq:group_der_taylor2}
    \! f(e^{i \varepsilon T^a} U) &\approx f(U) \! + \! i \epsilon \left(  T^a_{ik} U_{kj} \frac{\partial f}{\partial U_{ij}} \! - \!  T^a_{ki}  \frac{\partial f}{\partial \bar U_{ij}} (U^\dagger)_{jk} \right) \nonumber \\
    &= f(U) \! + \! i \epsilon \,\mathrm{Tr} \!\left[ T^a \!\left(U \frac{\partial f}{\partial U} \! - \! \frac{\partial f}{\partial U^\dagger} U^\dagger \right) \!\right]\!,
\end{align}
where we have used the following shorthands for complex matrix derivatives
\begin{align} \label{eq:matrix_derivatives}
    \left( \frac{\partial f}{\partial U} \right)_{ij} = \frac{\partial f}{\partial U_{ji}}, \quad 
    \left( \frac{\partial f}{\partial U^\dagger} \right)_{ij} = \frac{\partial f}{\partial \bar U_{ij}}.
\end{align}
Direct comparison with Eq.~\eqref{eq:group_der_taylor1} yields
\begin{align} \label{eq:group_der_matrix}
    D^a f(U) = i \mathrm{Tr} \left[ T^a \left(U \frac{\partial f}{\partial U} - \frac{\partial f}{\partial U^\dagger} U^\dagger \right) \right].
\end{align}
Analogously, we may replace the generator $T^a$ with an arbitrary algebra element $\hat K = K^a T^a$ as a linear combination of generators. Doing so yields a generalization of Eq.~\eqref{eq:group_der_taylor1},
\begin{align} \label{eq:exp_group_der_trick}
    f(e^{i \epsilon \hat K}U) = f(U) + \epsilon K^a D^a f(U) + \mathcal{O}(\epsilon^2).
\end{align}

\section{SDEs on non-abelian Lie groups} \label{app:stratonovich}

In this Appendix we clarify some properties of Stratonovich SDEs on non-abelian Lie groups and derive a few useful relations. We first start with a Stratonovich SDE for a single real variable $x^{(t)}$
\begin{align} \label{eq:app_start_sde}
    \dd x^{(t)} = k(x^{(t)}) \, \dd t + \sigma(x^{(t)}) \circ \dd W^{(t)},
\end{align}
where $k$ is the drift and $\sigma$ is the diffusion coefficient. The former is assumed to be continuous and at least differentiable, whereas the latter is twice differentiable.

Given a discretization of the diffusion time parameter $t$ with spacing $\epsilon$, the Stratonovich term approximated using the mid-point rule is
\begin{align} \label{eq:app_strat_term_mid}
     \sigma(x^{(t)}) \circ \dd W^{(t)} \approx \frac{\sigma(x^{(t+\epsilon)}) \! + \! \sigma(x^{(t)}) }{2} (W^{t+\epsilon} \! - \! W^t),
\end{align}
whereas the Itô product approximates at the base-point
\begin{align}
    \sigma(x^{(t)}) \cdot \dd W^{(t)} \approx \sigma(x^{(t)}) (W^{t+\epsilon} - W^t).
\end{align}
The difference between these two definitions is due to the fact that the increment of the Wiener process can be written as
\begin{align} \label{eq:app_noise_term}
    W^{t+\epsilon} - W^t = \sqrt{\epsilon} \, \eta, 
\end{align}
where $\eta$ is a random Gaussian noise with unit variance and zero mean, $\langle \eta \rangle_\eta = 0$ and $\langle \eta^2 \rangle_\eta = 1$. The presence of the $\sqrt{\epsilon}$ factor changes the usual power counting compared to discretization of ODEs.

Using Eq.~\eqref{eq:app_strat_term_mid}, the finite increment $\delta x = x^{(t+\epsilon)} - x^{(t)}$ can be written as
\begin{align}
    \delta x = \sqrt{\epsilon} \, \bar \sigma \, \eta + \epsilon \, k + \mathcal{O}(\epsilon^{3/2}),
\end{align}
where $\bar \sigma = (\sigma(x^{(t+\epsilon)}) + \sigma(x^{(t)})) / 2$ and the drift is evaluated at $x^{(t)}$. 

Given a twice differentiable function $f(x)$, we can consider the finite increment due to the SDE
\begin{align}
    \!\! \delta f &= f(x^{(t+\epsilon)}) - f(x^{(t)}) \nonumber \\
    &\approx \partial_x f \, \delta x + \frac{1}{2} \partial_x^2 f \, \delta x^2 +\mathcal{O}(\epsilon^{3/2}) \nonumber \\
    &\approx \sqrt \epsilon \partial_x f \, \bar \sigma  \eta + \epsilon \big[ \partial_x f \, k + \frac{1}{2} \partial_x^2 f \sigma^2 \eta^2 \big] +\mathcal{O}(\epsilon^{3/2}).
\end{align}
Rewriting the first term to consistently approximate around the mid-point,
\begin{align}
    \partial_x f \approx \overline{\partial_x f} -\frac{1}{2} \sqrt{\epsilon} \, {\partial_x^2 f} \, {\sigma} \, \eta + \mathcal{O}(\epsilon),
\end{align}
we find
\begin{align}
    \delta f 
    &\approx \sqrt \epsilon \,\, \overline{\partial_x f \, \sigma} \,   \eta + \epsilon \, \partial_x f \,  k +\mathcal{O}(\epsilon^{3/2}).
\end{align}
Comparing the above increment with $\delta x$ and Eq.~\eqref{eq:app_start_sde}, we recover the Stratonovich SDE for $f$,
\begin{align} \label{eq:app_strat_chain_rule}
    \dd f^{(t)} = \partial_x f(x^{(t)}) \, \circ \, \dd x^{(t)}.
\end{align}
This is the chain rule for Stratonovich SDEs.

On non-abelian Lie groups like SU$(N_c)$ and U$(N_c)$, we consider diffusion processes of the form
\begin{align} \label{eq:strat_sde_nonabelian}
    \dd U^{(t)} = i T^a \, \dd \Theta^a \circ U^{(t)},
\end{align}
where
\begin{align}
    \dd \Theta^a  = K^a(U^{(t)}) \, \dd t + g(t) \, \dd W^{(t), a}.
\end{align}
The process is group-preserving in the sense that $U^{(t)}$ remains unitary for all diffusion times $t$ and satisfies $\det U^{(t)} = 1$ in the case of SU$(N_c)$.

We seek to generalize the chain rule in Eq.~\eqref{eq:app_strat_chain_rule} for the non-abelian process. We consider a differentiable function $f(U)$ on the group evaluated for the process $U^{(t)}$. A straightforward application of the usual Stratonovich chain rule reads
\begin{align} \label{eq:df_nonabelian_step1}
    \dd f(U^{(t)}) = \frac{\partial f}{\partial U_{ij}} \circ \dd U^{(t)}_{ij} + \frac{\partial f}{\partial \bar U_{ij}} \circ \dd \bar U^{(t)}_{ij},
\end{align}
where $U^{(t)}_{ij}$ are the matrix elements of $U^{(t)}$. Because the process is group-preserving, $\dd U^{(t)}_{ij}$ and $\dd \bar U^{(t)}_{ij}$ are not independent. In terms of matrix elements, we find
\begin{align}
    \dd U^{(t)}_{ij} &= +i  \dd \Theta^a \circ (T^a_{ik} U^{(t)}_{kj}), \\
    \dd \bar U^{(t)}_{ij} &= -i  \dd\Theta^a \circ ( U^{\dagger (t)}_{jk} T^a_{ki}).
\end{align}
Inserting this into Eq.~\eqref{eq:df_nonabelian_step1} yields
\begin{align}
    \dd f(U^{(t)}) = i \, \mathrm{Tr}\left[ T^a \left( U \frac{\partial f}{\partial U} -\frac{\partial f}{\partial U^\dagger} U^\dagger \right)\right] \circ \dd \Theta^a,
\end{align}
where we have used the matrix derivatives defined in Eq.~\eqref{eq:matrix_derivatives}. Using the identity of Eq.~\eqref{eq:group_der_matrix}, we immediately obtain the chain rule
\begin{align} \label{eq:strat_sde_nonabelian_chain_rule}
    \dd f(U^{(t)}) = D^a f(U^{(t)}) \circ \dd \Theta^a. 
\end{align}

It is then also easy to see that the non-abelian process is indeed preserving unitarity by, e.g., considering the function
\begin{align}
    f(U) = \mathrm{Tr}[U U^\dagger],
\end{align}
which satisfies $D^a f(U) = 0$ and thus $\dd f = 0$.

Finally, the process in Eq.~\eqref{eq:strat_sde_nonabelian} is easily simulated numerically with time-step $\epsilon$ using the exponential update 
\begin{align} \label{eq:strat_sde_nonabelian_update_appendix}
    U^{(t+\epsilon)} = \exp\!\left( i T^a \!\!\left[ \epsilon \, K^a(U^{(t)}) \! + \! \sqrt{\epsilon} \, g(t) \, \eta^a \right] \right) U^{(t)}, 
\end{align}
where $\eta^a$ are independent noise terms with $\langle \eta^a \rangle = 0$ and $\langle \eta^a \eta^b \rangle = \delta^{ab}$. 

To show that the update scheme is indeed correct, we expand up to order $\epsilon^{3/2}$, dropping arguments for brevity and using $\hat K = K^a T^a$ and $\hat \eta = \eta^a T^a$ evaluated at time $t$,
\begin{align} \label{eq:exp_update_expansion}
    U' = (\mathds{1} + i \sqrt{\epsilon} g \, \hat \eta + i  \epsilon \hat K - \frac{1}{2} \epsilon \, g^2 \, \hat{\eta}^2) \, U + \mathcal{O}(\epsilon^{3/2}),
\end{align}
where $U = U^{(t)}$ and $U' = U^{(t+\epsilon)}$ (similarly for $g$ and $g'$).

We compare this to a direct discretization of Eq.~\eqref{eq:strat_sde_nonabelian} using the mid-point rule
\begin{align}
    \dd U &\approx \delta U = U' - U \nonumber \\
    &\approx i \sqrt{\epsilon} \, \hat \eta  \, \frac{g' U' + g U}{2} + i \epsilon \hat K \, U  + \mathcal{O}( \epsilon^{3/2}).
\end{align}
Using the leading order expansion (with $g' = g + \mathcal{O}(\epsilon)$)
\begin{align}
    g' U' = g (\mathds{1} +i \sqrt{\epsilon} \,g \, \hat \eta ) U + \mathcal{O}(\epsilon),
\end{align}
we obtain
\begin{align}
    \delta U &=  i \sqrt{\epsilon} \, g \, \hat \eta \,  U   + i \epsilon \hat K \, U - \frac{1}{2}  \epsilon  \, g^2 \, \hat \eta^2 \, U +  \mathcal{O}( \epsilon^{3/2}).
\end{align}
We see that Eq.~\eqref{eq:exp_update_expansion} and the above expression agree up to order $\epsilon^{3/2}$, which proves that the update scheme is correct.

\section{Analytical solution in two dimensions}
\label{app:analytical_solution}

In this Appendix, we provide the analytical solution for the forward diffusion process for U($N_c$) gauge groups in two dimensions, following Section~\ref{sec: analytical_diffusion}. We note that the results are applicable for general compact Lie groups. 
As observables, we consider rectangular Wilson loops and the topological charge susceptibility. We also present the Wilson loops in SU($N_c$); the topological charge susceptibility vanishes in that case. The analytical evolution can be compared with the numerically computed evolution to verify the implementation of the forward process. 

The partition function, with the Wilson action \eqref{eq:wilson_action} but ignoring the constant contribution, so we can write $S_E(\mathbf U)=\sum_P S_P(U_P)$,
takes the standard form in the character expansion \cite{Montvay:1994cy}
\begin{align}
\label{eq:expansionZ}
	\mathcal{Z} 
	= \int \prod_\ell \DD U_\ell\, \prod_P \sum_{r_P} d_{r_P} a_{r_P}(\beta)\,\chi_{r_P}(U_P).
\end{align}
Here the Boltzmann weight is expanded in terms of characters, where the summation is over all irreducible representations $r$ of U($N_c$) with dimension $d_r$ and coefficients
\begin{equation} \label{eq:expansion_coeffs}
	a_r(\beta ) = \frac{1}{d_r} \int \DD U_P \,  e^{-S_P(U_P)} \chi^{\ast}_r(U_P).
\end{equation}
On a two-dimensional lattice of genus $g$ and volume $V$ (and hence $V$ plaquettes), the partition function is evaluated exactly \cite{Migdal:1975zg,Rusakov:1990rs} as
\begin{align}
	\label{eq: Partition_repsum}
	\mathcal{Z} = \sum_{r} \left[ a_{r}(\beta) \right]^{V} d_r^{2-2g}.
\end{align}
For the following, we will consider periodic boundary conditions, therefore a torus with genus $g=1$.

Let us define the weight of each representation $r$

\begin{equation}
	w_r(\beta) = \frac{\left[ a_{r}(\beta) \right]^{V}}{\sum_{s} \left[ a_{s}(\beta) \right]^{V}}.
\end{equation}

Evaluating the average plaquette is straightforward

\begin{align}
	\langle W_{1 \times 1}[U]  \rangle & = \frac{1}{V}\frac{\partial \log \mathcal{Z} }{\partial \beta } \nonumber \\
	&= \sum_r w_{r}(\beta) \frac{\partial}{\partial \beta } \log a_{r}(\beta).
    \label{eq: wilson_loop_incomplete}
\end{align}

The derivative of the coefficients w.r.t.\ $\beta$ can be simplified using the character identity

\begin{equation}
    \chi_r(\mathbf{U}) \chi_s(\mathbf{U}) = \sum_\lambda \mathrm{N^\lambda_{r,s}}\; \chi_\lambda(\mathbf{U}),
\end{equation}
where $\mathrm{N^\lambda_{r,s}}$ are the Littlewood-Richardson coefficients \cite{FultonHarris:RepresentationTheory}, which count the multiplicity of $\lambda$ in the tensor product $\rm r \otimes s$ of representations, given its decomposition into a direct sum of irreducible representations $\{\lambda\}$,

\begin{equation}\label{eq: irrep_decomposition}
    \rm r \otimes s = \bigoplus_\lambda  \mathrm{N^\lambda_{r, s}} \lambda.
\end{equation}
Using the following relation:
   \begin{align}
        \partial_\beta a_r(\beta) 
        &=  \frac{1}{d_r} \int \DD U \frac{\partial}{\partial \beta }  e^{-S(U)} \chi^{\ast}_r(U) \nonumber\\
        &= \frac{1}{2 d_r N_c} \sum_s d_s a_s \int \DD U \left(\chi_f+\chi_{\bar{f}}\right) \chi^{\ast}_r\chi_s \nonumber \\
        &= \frac{1}{2d_r N_c} \sum_s d_s a_s(\beta) \mathrm{(N^r_{s,f} + N^r_{s,\bar{f}})}.
    \end{align}
The coefficient $\mathrm{N^r_{s,f}} = 1$ if the Young diagram of representation $\rm r$ is that of representation $\rm s$ with a box added, else $0$.
Using charge-conjugation symmetry $a_{\bar{s}} = a_s$, $d_{\bar{s}} = d_s$, $\mathrm{N^r_{s,\bar{f}}} = \mathrm{N^{\bar{r}}_{\bar{s}, f}}$, the sums of the two terms are symmetric under a relabelling ($\rm r \to \bar{r}$, $\rm s \to \bar{s}$) and so,
Eq.~\eqref{eq: wilson_loop_incomplete} becomes
    \begin{equation}
        \langle W_{1 \times 1} \rangle =
		\frac{1}{N_c}\sum_r w_{r}(\beta) \sum_{s} \mathrm{N^r_{s, f}}\frac{d_s a_s(\beta)}{d_r a_r(\beta)}.
    \end{equation}

The Wilson loop expectation values extend to general rectangular loops $\Gamma = R\times T$. Cutting the torus along $\Gamma $ leaves a disc of area $A$ and a punctured torus of area $V-A$, which share the boundary $\Gamma$. A surface of genus $g$ with one boundary has the partition function
\begin{equation}
    Z_{g}(A;U_\Gamma) = \sum_r d_r^{\,1-2g}\, a_r^{A}\,\chi_r(U_\Gamma).
\end{equation}
Gluing the two along $\Gamma$ and integrating over the common boundary gives 
\begin{equation}
    \langle W_{R \times T} \rangle = \frac{1}{N_c} \sum_{r} w_r(\beta) \sum_s \mathrm{N^r_{s, f}} \frac{d_s}{d_r} \left[\frac{a_s(\beta)}{a_r(\beta)}\right]^{RT}.
\end{equation}

For a loop that is small compared to the lattice, $RT \ll L^2$, the leading order dominates and
\begin{equation}
    \langle W^{(r)}_{R\times T}\rangle = \left[\frac{a_r(\beta)}{a_0(\beta)}\right]^{RT}.
\end{equation}

The above process can be adapted to topological quantities, as in \cite{Bonati:2019olo}.
Introducing a $\theta$-dependent partition function,
\begin{align}
	\mathcal{Z}_\theta &= \int \prod_\ell \DD U_\ell\, e^{-S[U] + i\theta Q},
\end{align}
leads to the same form as Eq.~\eqref{eq: Partition_repsum}, now with $\theta$-dependent coefficients
\begin{equation}\label{eq:expansion_coeffs_w_theta}
	\gamma_r(\beta , \theta ) = \frac{1}{d_r} \int \DD U_P\; \chi^{\ast}_r(U_P)\, e^{-S_P(U_P) + i \theta \, q_P},
\end{equation}
which reduce to $a_r(\beta)$ at $\theta=0$. Writing $\gamma_r'$ and $\gamma_r''$ for the first and second $\theta$-derivatives at $\theta=0$, the charge and its second moment are
\begin{align}
	\langle Q \rangle &= -i\, \frac{\partial \log\mathcal{Z} _\theta}{\partial \theta } \Big|_{\theta =0} = -iV\sum_{r} w_r \frac{\gamma_r^{\prime}}{\gamma_r} = 0, \\
	\langle Q^2 \rangle &= -\sum_{r} w_r \left[V\frac{\gamma_r''}{\gamma_r} + V(V-1)\left( \frac{\gamma_r'}{\gamma_r} \right)^2\right]. \label{eq:gamma_theta_coeffs}
\end{align}

In two dimensions, the topological charge is set by the U(1) factor of U($N_c$), through the phase of the determinant, $\det U_P =  e^{i \phi_P }$. The determinant is a one-dimensional character; setting $\lambda_i = k$ for all $i$ in Weyl's character formula,
\begin{align}
    \chi_\lambda(U) = \frac{\det_{1 \le i,j \le N_c}\left[z_j^{\lambda_i + N_c - i}\right]}{\det_{1 \le i,j \le N_c}\left[z_j^{N_c - i}\right]},
\end{align}
gives $\chi_{(k,\dots,k)}(U) = (\det U)^k$ with $d_{(k,\dots,k)}=1$.

The plaquette charge $q_P = \phi_P/(2\pi)$ is a sawtooth function in $\phi_P$, with the Fourier series
    \begin{equation}\label{eq:charge_expansion}
        q_P = \sum_{k\neq 0} c_k\, (\det U_P)^k, \qquad c_k =i\,\frac{(-1)^k}{2\pi k}.
    \end{equation}
Multiplying by $(\det U_P)^k$ shifts every weight of a representation, $r \to r+k$, so a $\det^k$-insertion changes the coefficients $a_r \to a_{r-k}$. 
The $\theta$-derivatives of the coefficients are single-plaquette integrals with $q_P$ inserted,
    \begin{align}
    	\gamma_r'(\beta) &= \frac{i}{d_r} \int \DD U_P\; q_P\,\chi^{\ast}_r(U_P)\, e^{-S_P(U_P)},\\
    	\gamma_r''(\beta) &= -\frac{1}{d_r} \int \DD U_P\; q_P^2 \,\chi^{\ast}_r(U_P)\, e^{-S_P(U_P)}.
    \end{align}
Expanding $q$ and $q^2$ in the characters $(\det U_P)^k$ as in Eq.~\eqref{eq:charge_expansion}, and using that each insertion shifts $a_r \to a_{r-k}$, these become
    \begin{align}
        \gamma_r' = i \sum_{k\neq 0} c_k\, a_{r-k}, \quad \gamma_r'' = -\sum_{m\in \mathbb{Z}} s_m\, a_{r-m}, 
    \end{align}
where $s_m=\sum_{k+k'=m} c_k c_{k'}$ are the Fourier coefficients of $q^2$, namely $s_0 = \tfrac1{12}$ and $s_{m\neq0}=\tfrac{(-1)^m}{2\pi^2m^2}$. Since $\gamma_r(\beta, 0) = a_r$, dividing by it gives
    \begin{equation}
        \frac{\gamma_r'}{\gamma_r} = i f_r, \quad \frac{\gamma_r''}{\gamma_r} = -S_r,
    \end{equation}
with 
    \begin{align}
        f_r &= \frac{i}{2\pi} \sum_{k\neq 0} \frac{(-1)^k}{k}\, \frac{a_{r-k}}{a_r}, \\
        S_r &= \frac{1}{12} + \frac{1}{2\pi^2}\sum_{m\neq 0} \frac{(-1)^m}{m^2}\, \frac{a_{r-m}}{a_r}.
    \end{align}
Inserting these into Eq.~\eqref{eq:gamma_theta_coeffs} yields
    \begin{equation}
        \langle Q^2\rangle_0 = V\sum_r w_r\left[S_r + (V-1) f_r^2\right].
    \end{equation}
    
The same result follows directly from the plaquette correlators. For distinct plaquettes the $\det^k$ and $\det^{k'}$ insertions factorise at fixed $r$,
\begin{align}
    \langle q_P q_{P'}\rangle_0 &= \sum_r w_r f_r^2,
\end{align}
while for $P=P'$ the two insertions merge into a single $\det^{k+k'}$, and the double sum reduces to the Fourier series of $q^2$,
\begin{align}
    \langle q_P^2\rangle_0 = \sum_r w_r S_r.
\end{align}
Then $\langle Q^2\rangle_0 = V\langle q_P^2\rangle_0 + V(V-1) \langle q_P q_{P'}\rangle_0$ reproduces the expression above.

Under diffusion, there are three cases that evolve differently, because the U(1) charges of the two plaquettes combine differently on their links. Every link is an eigenfunction of the Laplacian with damping $e^{-\frac{\tau}{2}\frac{N_c}{2} n^2}$, where $n$ is the net charge on that link; a shared link carries $k-k'$, an unshared one carries the charge of its plaquette. Summing over the four links of each plaquette gives the total exponents

\begin{equation}
	C_{kk'} =
	\begin{cases}
		(k+k')^2 & P = P', \\
		k^2 + k'^2 - \tfrac{1}{2}kk' & P \neq P' \text{ adjacent}, \\
		k^2 + k'^2 & P \neq P' \text{ non-adjacent},
	\end{cases}
\end{equation}
and each term is damped by $e^{-\tau\, N_c C_{kk'}}$.
A lattice with $V$ plaquettes has $V(V-1)$ ordered pairs with $P\neq P'$, of which $4V$ are adjacent, hence
\begin{align}
    \langle Q^2\rangle_\tau &= V \sum_r w_r \left[
    S_r(\tau) + 4\, A_r(\tau) + (V-5)\, B_r(\tau) \right]
\end{align}
with
\begin{align}
	S_r(\tau) &= \frac{1}{12} + \frac{1}{2\pi^2}\sum_{k\neq 0}\frac{(-1)^k}{k^2}\,\frac{a_{r-k}}{a_r}\,e^{-\tau N_c k^2}, \\
	A_r(\tau) &= -\frac{1}{4\pi^2}\!\!\sum_{k,k'\neq 0}\!\!\!\frac{(\!-1)^{k+k'}}{kk'}\,\frac{a_{r-k}\,a_{r-k'}}{a_r^2}\,e^{-\tau N_c (k^2 \! +  k'^2 \! - \! \frac{kk'}{2})}, \\
	B_r(\tau) &= -\frac{1}{4\pi^2}\sum_{k,k'\neq 0}\frac{(-1)^{k+k'}}{kk'}\,\frac{a_{r-k}\,a_{r-k'}}{a_r^2}\,e^{-\tau N_c (k^2 + k'^2)}.
\end{align}
As $\tau\to\infty$, only the constant term survives, so $\chi_{\rm top} \to 1/12$.

For $U(N_c)$, an irreducible representation is uniquely specified by its highest weight
\begin{equation}
    \lambda = (\lambda_1, \ldots, \lambda_{N_c}), \qquad \lambda_1 \geq \cdots \geq\lambda_{N_c},
\end{equation}
where $\{\lambda_i\}^{N_c}_{i=1}$ are integers. For representations described by Young tableaux, they are non-negative, $\lambda_i\geq 0$, giving the length of the $i$-th row of the tableau; see e.g.\ Refs.~\cite{FultonHarris:RepresentationTheory, Cvitanovic}.
The coefficients in expansion \eqref{eq:expansionZ} then read
\be
a_{\lambda}(\beta) = \frac{1}{d_\lambda} \det_{1 \le i,j \le N_c}\left[ I_{\lambda_i - i + j}\!\left(\frac{\beta}{N_c}\right) \right],
\ee
where $I_\lambda(z)$ are modified Bessel functions of the first kind,
and the dimensions are
\be
d_{\lambda} = \prod_{i<j} \frac{\lambda_i - \lambda_j + j - i}{j - i}.
\ee
The quadratic Casimir reads explicitly
\be
C_2(\lambda) = \half \sum_{i=1}^{N_c} \lambda_i \left( \lambda_i + N_c + 1 - 2i \right).
\ee

The extension to SU($N_c$) is obtained by restricting the U($N_c$) representations, with the highest weights $\lambda=\{\lambda_i\}_{i=1}^{N_c}$, under the equivalence relation $\lambda_i \sim \lambda_i + k $ for all $i$ and $k\in \mathbb{Z}$. Weights that differ by a common integer shift restrict to the same SU($N_c$) representation which we label by the representative with vanishing last entry,
\begin{equation}
    [\lambda] = \{\lambda_i - \lambda_{N_c}\}_{i=1}^{N_c}.
\end{equation}

Since $\det U = 1$ on SU($N_c$), the characters $\chi_{\lambda + k}$ all coincide with $\chi_{[\lambda]}$, and the SU($N_c$) coefficients are the sums of the U($N_c$) coefficients over each equivalence class,
\begin{equation}
    a_{[\lambda]}^{\mathrm{SU}(N_c)}(\beta) = \sum_{k\in \mathbb{Z}} a_{\lambda + k}^{\mathrm{U}(N_c)}(\beta).
\end{equation}
The dimensions remain unchanged by the shift.

The Casimirs of $\mathrm{U}(N_c)$ and $\mathrm{SU}(N_c)$ differ by the contribution of the $\mathfrak{u}(1)$ factor ($\mathfrak{u}(N_c) = \mathfrak{su}(N_c) \oplus \mathfrak{u}(1)$), which must be subtracted,

\begin{equation}
    C_2^{\rm SU(N_c)}([\lambda]) = C_2^{\rm U(N_c)}(\lambda) - \frac{q(\lambda)^2}{2N_c},\; q(\lambda) = \sum_{i=1}^{N_c}\lambda_i.
\end{equation}
As a check, the fundamental representation $\lambda = (1, 0, \dots, 0)$ gives the standard value
\begin{equation}
    C_2^{\mathrm{SU}(N_c)}([\mathrm{f}]) = \frac{N_c}{2} - \frac{1}{2N_c} = \frac{N_c^2 - 1}{2N_c}.
\end{equation}
Similarly, $\lambda = (k, \dots, k)$ gives $C_2^{\mathrm{SU}(N_c)}([\rm{\lambda}])=0$, consistent with $\det U^k$ being trivial on SU($N_c$).

\section{On the effective time-dependent action} \label{app:eff_action}

In this Appendix, we explicitly derive the time dependence of the plaquette correlators (see Eq.~\eqref{eq:plaq_correlators}) under the forward process. The diffusion process acts on each link through the semigroup \eqref{eq: semigroup_on_link}. However, by Eq.~\eqref{eq:expansionZ}, every link is shared by two plaquettes, so the action on a link couples a pair of plaquettes, and their character product is in general not an eigenfunction of the semigroup.

Consider a link $U_\ell$ shared by two plaquettes, $U_P = A\, U_\ell$ and $U_{P'} = U_\ell^{-1} B$. The product of their characters is a trace over the tensor product representation $r \otimes \bar{s}$, which we decompose into irreducible representations,
    \begin{align}
        \chi_r(U_P) \chi_s(U_{P'}) 
        &= \Tr_{r \otimes \bar{s}}\left[U_{P} \otimes U_{P'}\right] \nonumber\\
        &= \sum_{\lambda \subset \mathrm{r} \otimes \mathrm{\bar{s}}} \Tr_{\lambda}\left[P_\lambda(U_{P} \otimes U_{P'})\right],
    \end{align}
where $P_\lambda$ projects onto the subspace of representation $\lambda$ and $\sum_\lambda P_\lambda = 1$. 
The Laplacian on the shared link acts on both factors,
\begin{equation}
    \hat\Delta_\ell = -C_2(r) - C_2(s) + 2\, \Omega_{rs},
    \quad \Omega_{rs} = \sum_a T^a_r \otimes T_s^a,
\end{equation}
and in the subrepresentation $\lambda$ the cross term is fixed by the Casimirs,
\begin{equation}
    \Omega_{rs}\big|_\lambda = - \half \left[C_2(\lambda) - C_2(r) - C_2(s)\right].
\end{equation}
Hence,

\begin{equation}
    e^{\frac{\tau}{2}\Delta_\ell} \chi_r(U_P)\chi_s(U_{P'}) = \sum_\lambda e^{-\frac{\tau}{2}C_2(\lambda)} \Tr_\lambda\left[P_\lambda(U_P \otimes U_{P'})\right].
    \label{eq: pair_irrep_sum}
\end{equation}

Integrating over the shared link $U_\ell$ keeps only the singlet channel, which has $C_2=0$ and therefore does not evolve. Links enclosed by a loop $\Gamma$ thus do not contribute, and only the links on $\Gamma$ evolve. For a non-intersecting loop $\Gamma$, we find
\begin{align}
	\prod_{\ell \in \Gamma} e^{\frac{\tau}{2}\hat\Delta_\ell}  \chi_{r}(U_\Gamma) = e^{-\frac{\tau}{2} \mathcal P(\Gamma)\, C_2(r)}\, \chi_{r}(U_\Gamma),
\end{align}
where $\mathcal  P(\Gamma)$ is the perimeter of $\Gamma$.

Expectation values then evolve as
\begin{equation}
    \langle O\rangle_\tau
        = \langle e^{\frac{\tau}{2} \hat\Delta}\, O \rangle_{0}.
    \label{eq: observable_time_evolution}
\end{equation}
and for a rectangular loop, with $\mathcal P(\Gamma)=2(R+T)$,
\begin{equation}
    \langle W^{(r)}_{R\times T}\rangle_\tau = e^{-\tau (R+T)\, C_2(r)} \, \langle W^{(r)}_{R\times T}\rangle_0.
\end{equation}
Let $O_\pm = \frac{1}{N_c^2} \Tr U_P \Tr U_{P'}^{\pm 1}$ for two adjacent plaquettes sharing a link, and $O_P = \frac{1}{N_c} \Tr U_P$. 

By Eqs.~\eqref{eq: pair_irrep_sum} and \eqref{eq: observable_time_evolution},
\begin{equation}
    e^{\frac{\tau}{2} \Delta_\ell} O_{\pm} = \frac{1}{N_c^2} \sum_\lambda e^{-\frac{\tau}{2}C_2(\lambda)} \Tr\left[P_\lambda (U_{P} \otimes U_{P'}^{\pm 1})\right].
\end{equation}

The six unshared links are in the fundamental representation and contribute $e^{-\frac{\tau}{2} \cdot 6 \cdot \frac{N_c}{2}} = e^{-\frac{3}{2} N_c \tau}$. The shared link carries $\mathbf{f} \otimes \mathbf{\bar{f}} = \mathbf{1} \oplus \mathbf{Adj}$ for $O_+$ and $\mathbf{f} \otimes \mathbf{f} = \mathbf{Sym^2f} \oplus \mathbf{\bigwedge^2 f}$ for $O_-$. Each channel is damped accordingly by its Casimir. Hence, in the large volume limit,

\begin{align}
    & \langle O_+ \rangle_\tau = \frac{\langle O_P \rangle_0^2}{N_c^2} e^{-\frac{3}{2}N_c\tau} [1 +  (N_c^2 - 1) e^{-\frac{\tau}{2}N_c}], \\
    & \langle O_-\rangle_\tau = \frac{\langle O_P \rangle_0^2}{N_c^2} e^{-\frac{3}{2}N_c\tau} \\
    & \quad \times\left[\frac{N_c(N_c + 1)}{2} e^{-\frac{\tau}{2}(N_c+1)} + \frac{N_c(N_c - 1)}{2}e^{-\frac{\tau}{2}(N_c-1)}\right],  \nn \\
    & \langle O_P\rangle_\tau^2 = \langle O_P \rangle_0^2\; e^{-2 N_c\tau}.
\end{align} 
For small $\tau$, the difference increases linearly, 
\begin{equation}
  \langle O_+ \rangle_\tau -  \langle O_- \rangle_\tau = \frac{\langle O_P \rangle_0^2}{N_c^2} \left( N_c\tau + {\cal O}(\tau^2) \right). 
 \end{equation}

We can naturally extend this to arbitrary representations
\begin{align}
    \langle O_\pm^{(r)} \rangle_\tau &= \frac{\langle O^{(r)}_P\rangle_0^2}{d_r^2} e^{-3\, C_2(r)\,\tau} \sum_{\lambda} \mathrm{N}^\lambda_{r,s} d_\lambda\, e^{-\frac{\tau}{2}\, C_2(\lambda)}.
\end{align} 
where $\rm s=\bar{r}$ for $O_+^{(r)}$, and $\rm s=r$ for $O_-^{(r)}$.

For SU(2), the traced plaquette is real and $O_+=O_-$. Instead we consider the difference with the square of the expectation value of the traced plaquette, see Eq.~(\ref{eq:Opm2}), and find
\be
\bra O_\pm\ket_\tau - \bra O_P\ket^2_\tau =
\frac{\bra O_P\ket_0^2}{4} e^{-9\tau/4} \left[ 1+3 e^{-\tau} -4 e^{-3\tau/4} \right],
\ee
with, in the large volume limit, $\bra O_P\ket_0=e^{-\sigma}$.
At small $\tau$, the difference increases as $\tau^2$ instead of $\tau$.

\section{Exact Metropolis adjustment on compact groups} \label{app:metropolis}

In MAALA sampling, we propose updates of field configurations as
\begin{equation}\label{eq:app_langevin_step}
 U'_{x,\mu}=
\exp\!\left[ i \left( \epsilon \hat s_{\theta}(\mathbf{U})_{x,\mu}
+\sqrt{2\epsilon} \, \hat \eta_{x,\mu} \right) \right]\, U_{x,\mu},
\end{equation}
which we aim to accept or reject such that the samples along the Markov chain (asymptotically) follow the target probability $p_0(\mathbf{U})$ given by the Wilson action. As before, $\hat{\eta}_{x,\mu}=T^a\eta^a_{x,\mu}$, and we suppress time indices in the score for compactness. The standard Metropolis-Hastings algorithm leads to the acceptance probability presented in Eq.~\eqref{eq:metro_acc}, which is seemingly an approximation as one discards multiple wrappings. 
In this appendix, we show how we can reinterpret Eq.~\eqref{eq:metro_acc} as an exact acceptance probability in an extended sampling space. This reinterpretation is akin to HMC or the framework of Involutive MCMC \cite{pmlr-v119-neklyudov20a, cusumanotowner2020automatinginvolutivemcmcusing}.

The main idea is to extend the target space with an auxiliary variable (in our case $\eta=\{\eta^a_{x,\mu} | \forall a,x,\mu \}$ or equivalently $\hat\eta$) such that the target probability is now the joint probability 
\begin{equation}
    p(\mathbf{U},\hat\eta) = p_0(\mathbf{U}) q(\hat\eta).
\end{equation}
The auxiliary distribution $q$ is given as
\begin{equation}
    q(\hat\eta) = \prod_{a,x,\mu} \mathcal{N}(\eta^a_{x,\mu}|0,1).
\end{equation}
This is obviously very similar to HMC, where we introduce an auxiliary momentum variable for each degree of freedom in the theory, which follows a normal distribution. An additional ingredient is an update function $f(x) = x'$, which takes the combined state $x = (\mathbf{U}, \hat \eta)$ to a new state $x' = (\mathbf{U}', \hat \eta')$. This update function should leave the ``Hamiltonian'' $H(x) = - \ln p_0(\mathbf U) - \ln q(\hat \eta)$ approximately invariant, with a possibly small violation $\Delta H = H(x') - H(x)$. If $f$ is reversible and volume-preserving (as in the case of HMC, where we use solutions to Hamiltons equations), then the transition probability of the update is symmetric and the
new state $x'$ is accepted with the probability
\begin{align}
    p_{\text{accept}}
    &= \min \left\{ 1, e^{-\Delta H} \right\} = \min \left\{ 1, \frac{p_0(\mathbf{U}')}{p_0(\mathbf{U})}\frac{q(\hat \eta')}{q(\hat \eta)} \right\}.
\end{align}

Here, we use a discrete update proposal $f(x) = x'$ of the form
\begin{align}
    {U}'_{x,\mu} &= e^{i(\epsilon \hat{s}_{\theta}(\mathbf{U})_{x,\mu}+\sqrt{2\epsilon}\hat\eta_{x,\mu})}U_{x,\mu}, \\
    \hat\eta'_{x,\mu} &= - \hat\eta_{x,\mu} - \sqrt{\frac{\epsilon}{2}} \left( \hat{s}_{\theta}(\mathbf{U})_{x,\mu}+ \hat{s}_{\theta}(\mathbf{U'})_{x,\mu} \right).\label{eq:etaprime_update}
\end{align}
This update function has two important properties: 
firstly, it is an involution, i.e., $f(x)=f^{-1}(x) $ or $ f(f(x))=x$.  The involution property is analogous to reversibility in HMC and can be easily checked explicitly. We note that the discrete update has the same form as the one used in the main text based on the principal branch. As comparison with Eq.~\eqref{eq:metro_acc}, assume that 
\begin{align}
    \label{eq:app_eta}
    \hat \eta_{x,\mu} &= \frac{1}{i \sqrt{2 \epsilon}} \ln(U'_{x,\mu} {U}_{x,\mu}^{\dagger}) - \sqrt{\frac{\epsilon}{2}} \hat s_{\theta}(\mathbf{U})_{x,\mu},
\end{align}
then 
\begin{align}
    \hat\eta'_{x,\mu} &=  -\frac{1}{i \sqrt{2 \epsilon}} \ln(U'_{x,\mu} {U}_{x,\mu}^{\dagger}) + \sqrt{\frac{\epsilon}{2}} \hat s_{\theta}(\mathbf{U})_{x,\mu} \nonumber \\
    &\quad - \sqrt{\frac{\epsilon}{2}}\left( \hat{s}_{\theta}(\mathbf{U})_{x,\mu} + \hat{s}_{\theta}(\mathbf{U'})_{x,\mu} \right) \nonumber \\
    &= \frac{1}{i \sqrt{2 \epsilon}} \ln\left( (U'_{x,\mu} {U}_{x,\mu}^{\dagger})^{-1} \right) - \sqrt{\frac{\epsilon}{2}} \hat s_{\theta}(\mathbf{U'})_{x,\mu},
\end{align}
which is exactly Eq.~\eqref{eq:app_eta} with $\mathbf U$ and $\mathbf U'$ exchanged, as also defined in the main text. However, for exactness we stress that $\hat \eta$ and $\hat \eta'$ must not be computed from the matrix logarithm. Instead, $\hat \eta$ is simply stored when it is generated and $\hat \eta'$ is computed from Eq.~\eqref{eq:etaprime_update}.

Secondly, the function $f$ is also volume- or measure-preserving. To see this, we can factorize $f$ into simpler steps $g,h,k$ as
\begin{align}
f &= k \circ g \circ h \circ g, \\
g(\mathbf{U},\hat\eta) &= \Big(\mathbf{U},\ \hat\eta + \sqrt{\frac{\epsilon}{2}}\,\hat s_\theta(\mathbf{U})\Big), \\
h(\mathbf{U},\hat\eta) &= \left(\big\{e^{i\sqrt{2\epsilon}\,\hat\eta_{x,\mu}}\,U_{x,\mu}\big\},\ \hat\eta\right), \\
k(\mathbf{U},\hat\eta) &= (\mathbf{U},\ -\hat\eta).
\end{align}
Shifts (step $g$) and reflections (step $k$), for fixed $\mathbf U$, leave the measure of $\hat \eta$ invariant. Due to the left invariance of the Haar measure,  step $h$, at fixed $\hat \eta$, also leaves $\mathcal D U$ invariant. As a composition of measure-preserving steps, $f$ is measure-preserving as well.

With the target $p_0(\mathbf{U})$ given by the Wilson action, we get
\begin{align}
    \frac{p_0(\mathbf{U}')}{p_0(\mathbf{U})}\frac{q(\hat \eta')}{q(\hat \eta)} & =  \exp \Big[ S(\mathbf{U}) - S(\mathbf{U}') \nonumber \\
    &\quad+ \frac{1}{2}\sum_{a,x,\mu}\left( (\eta^a_{x,\mu})^2 - (\eta'^a_{x,\mu})^2 \right) \Big],
\end{align}
which can be easily calculated, and coincides with Eq.~\eqref{eq:metro_acc} ($\tr{\hat{\eta}^2} = \sum_a(\eta^a)^2/2$), if the principal branch of the matrix logarithm inverts the exponential. In our numerical experiments that is always the case due to the small stepsize $\epsilon$.

With these ingredients one can construct an MCMC chain which consists of the following steps: First, sample $\hat\eta$ from $q(\hat\eta)$, then apply the update $(\mathbf{U}',\hat\eta') = f(\mathbf{U},\hat\eta)$, and finally accept $\mathbf{U}'$ as a new configuration of the chain with probability $p_{\text{accept}}$.
This chain asymptotically samples $p(\mathbf{U},\hat\eta)$, and thus $p_0(\mathbf{U})=\exp(-S(\mathbf{U}))/\mathcal{Z}$, exactly, without the need to consider multiple wrappings of compact groups.

\section{Error estimates} \label{app:errors}
Given a set of field configurations, we calculate the standard error of the observables using the bootstrap method. For the results in Fig.~\ref{fig:beta_extrapolation}, we further combined the means and standard errors obtained from ten independently trained networks with identical hyperparameters using the DerSimonian-Laird analysis. Given some mean values $O_i$ and uncertainty estimates $\delta O_i$ ($i$ runs over the trained networks), one first calculates the weighted mean
\begin{align}
    \bar{O} = \frac{\sum_i w_i O_i}{\sum_i w_i}, \quad w_i=\frac{1}{\delta O_i^2}.
\end{align}
Further, we calculate
\begin{align}
    \tau^2 &= \max(0, \frac{Q-(N-1)}{C}), \\
    Q &= \sum_i w_i (O_i - \bar{O})^2, \\ 
    C &= \sum_i w_i - \frac{\sum_i w_i^2}{\sum_i w_i}, \\
    w^*_i &= \frac{1}{\delta O_i^2 + \tau^2},
\end{align}
which yields the adjusted weighted mean and the total random-effects uncertainty
\begin{align}
    O^* = \frac{\sum_i w^*_i O_i}{\sum_i w^*_i}, \quad \sigma^2_\mathrm{total} = \frac{1}{\sum_i w^*_i}.
\end{align}

\newpage

\bibliography{references}

\end{document}